\documentclass[useAMS,usenatbib]{mn2e}
\usepackage{graphicx}
\usepackage{epsfig}
\usepackage{graphicx}
\usepackage{epstopdf}
\usepackage{hyperref}
\usepackage{amsmath}
\usepackage{amsfonts}
\usepackage{times}
\usepackage{graphicx}
\usepackage{natbib}
\usepackage{nicefrac}

\renewcommand{\vec}[1]{\mbox{\boldmath $#1$}}

\def\i{{\rm i}}
\def\d{{\rm d}}
\def \Om  {{\it \Omega}}
\def \rin {r_{\rm in}}
\def \mur {r_{\rm in}}
\def\Rin{R_{\rm in}}
\def\Rout{R_{\rm out}}
\def \Rey {\ensuremath{\rm{Re}}}
\def \Remax {\ensuremath{\rm Re_{max}}}
\def \Ha {\ensuremath{\rm{Ha}}}

\def \Pm {\ensuremath{\rm{Pm}}}
\def \Rm {\ensuremath{\rm{Rm}}}
\def\etaT{\eta_{\rm T}}

\def\C{Chandrasekhar}
\def \Mm {\ensuremath{\rm{Mm}}}
\def \S  {\ensuremath{\rm{S}}}
\def \Lu  {\ensuremath{\rm{S}}}

\def\beg{\begin{equation}}
\def\ende{\end{equation}}
\newcommand{\gsim}{\lower.7ex\hbox{$\;\stackrel{\textstyle>}{\sim}\;$}}
\newcommand{\lsim}{\lower.7ex\hbox{$\;\stackrel{\textstyle<}{\sim}\;$}}
\renewcommand{\vec}[1]{\mbox{\boldmath $#1$}}
\def\curl{{\rm curl}} 
\def\Om{{\it \Omega}}

\def\A{{Alfv\'en}}

\def\ara\&a{ Ann. Rev. Astronomy Astrophysics}

\author[G. R\"udiger, M. Schultz]
{
G. R\"udiger$^{1,2}$\thanks{E-mail: GRuediger@aip.de},
M. Schultz$^1$, 
\\
$^1$Leibniz-Institut f\"ur Astrophysik Potsdam, An der Sternwarte 16, 14482 Potsdam, Germany\\
$^2$ University of Potsdam, Institute of Physics and Astronomy, Karl-Liebknecht-Str. 24-25, 14476 Potsdam, Germany
}

\title[Dynamo action   of magnetized    Taylor-Couette flows]{Large-scale dynamo action   of magnetized    Taylor-Couette flows}
\begin{document}
\date{Accepted . Received ; in original form }

\pagerange{\pageref{firstpage}--\pageref{lastpage}} \pubyear{2019}

\maketitle

\label{firstpage}

\begin{abstract}
A  conducting Taylor-Couette flow with quasi-Keplerian rotation law  
 containing a toroidal  magnetic field    serves as a mean-field dynamo model of the Tayler-Spruit-type. The  flows are unstable against nonaxisymmetric perturbations  which form   electromotive forces defining $\alpha$ effect and eddy diffusivity. 
If both degenerated modes with $m=\pm 1$ are excited with the same power then the global $\alpha$ effect vanishes and a  dynamo cannot work. 
It is shown, however,  that the Tayler instability     produces finite $\alpha$ effects  if  only an isolated  mode is considered but   
this intrinsic helicity of the single-mode  is  too low for an $\alpha^2$ dynamo.  Moreover, an $\alpha\Om$ dynamo model with quasi-Keplerian rotation requires   a minimum magnetic Reynolds number of rotation of ${\rm Rm}\simeq 2.000$ to work. 
 Whether it  really  works  depends on    assumptions about the turbulence energy.    
 For a steeper-than-quadratic   dependence of the  turbulence intensity  on the magnetic field, however,  dynamos are only excited   if the resulting magnetic eddy diffusivity approximates its microscopic value, $\eta_{\rm T}\simeq \eta$.  By basically   lower or  larger eddy  diffusivities  the  dynamo instability is suppressed.
\end{abstract}
\section{Introduction}
 A Taylor-Couette flow with a smooth  rotation profile beyond the Rayleigh limit becomes unstable if the magnetic background field is toroidal (or has a toroidal component) for relatively low  critical Reynolds numbers. If the magnetic field is strictly axial then the Reynolds numbers necessary for instability  are higher by several orders of magnitude if the magnetic Prandtl number $\Pm=\nu/\eta$ (with $\nu$ the molecular viscosity and  $\eta$ the molecular resistivity)
is much smaller than unity.  The reason is that the instabilities set in at  fixed Reynolds number for toroidal background fields and  at fixed magnetic Reynolds number for axial background fields which basically  differ for $\Pm\ll 1$.

\begin{figure*}
  \centerline{
  \hbox{
  \includegraphics[width=0.33\textwidth]{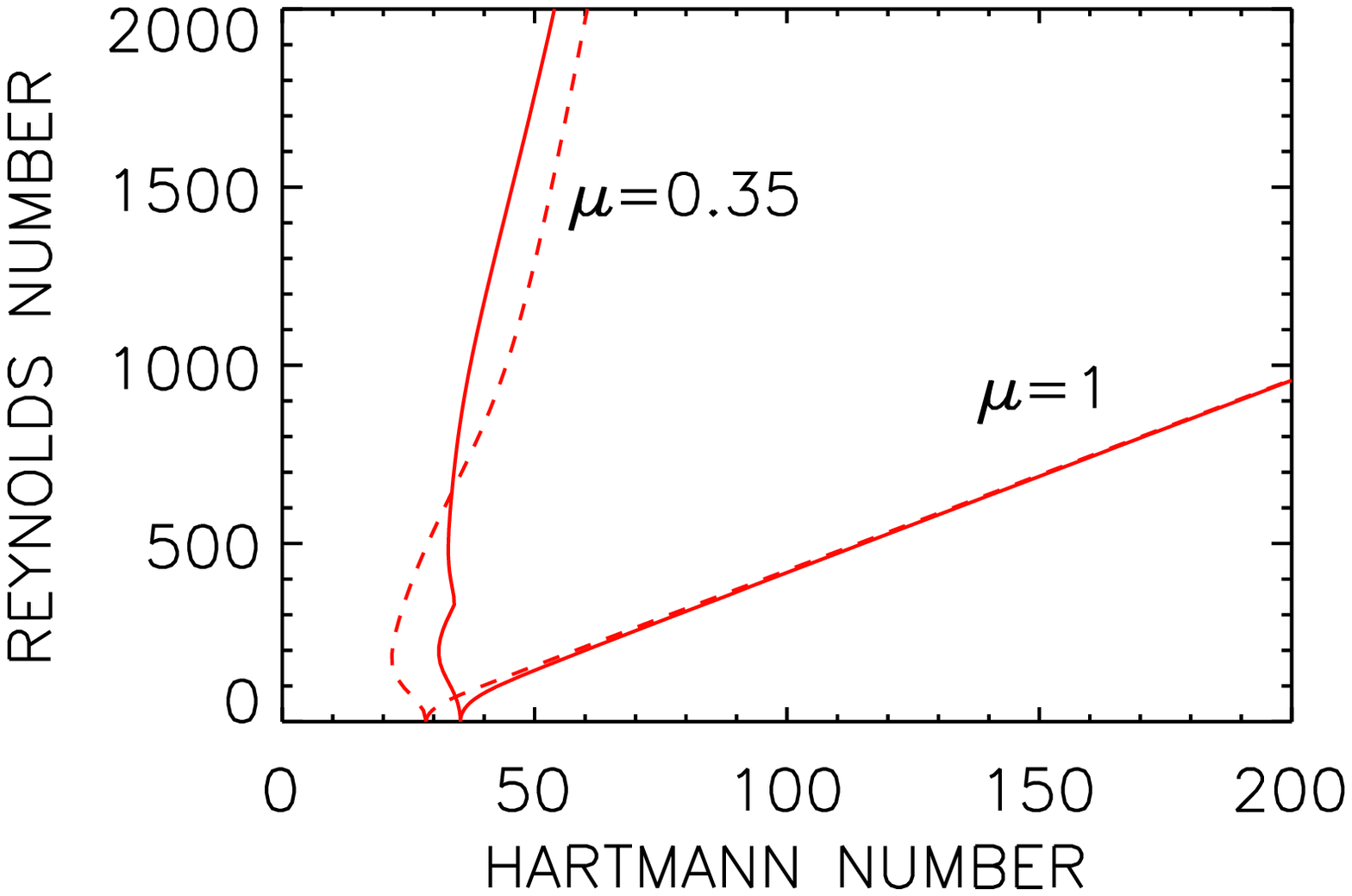}
  \includegraphics[width=0.33\textwidth]{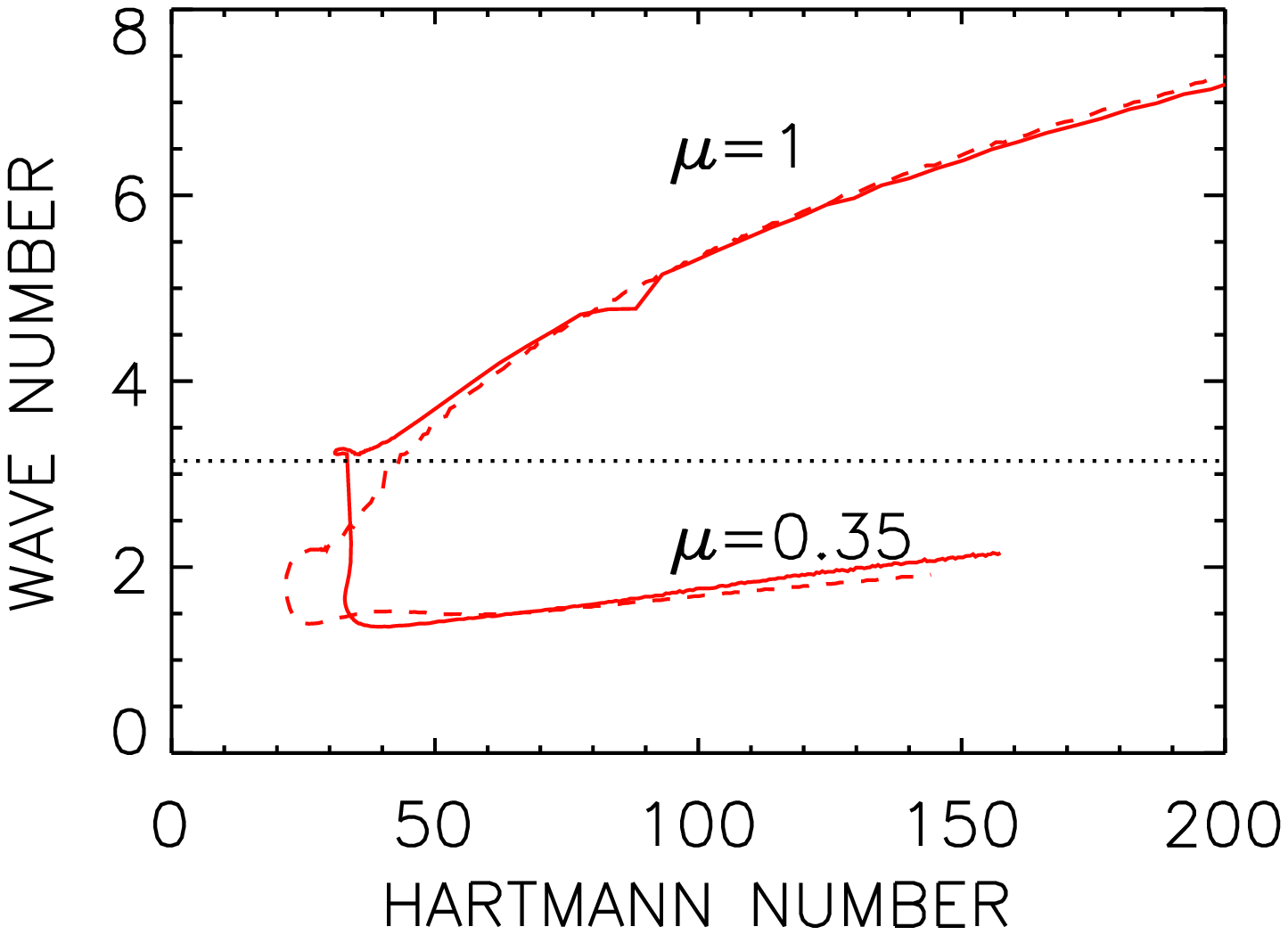}
  \includegraphics[width=0.33\textwidth]{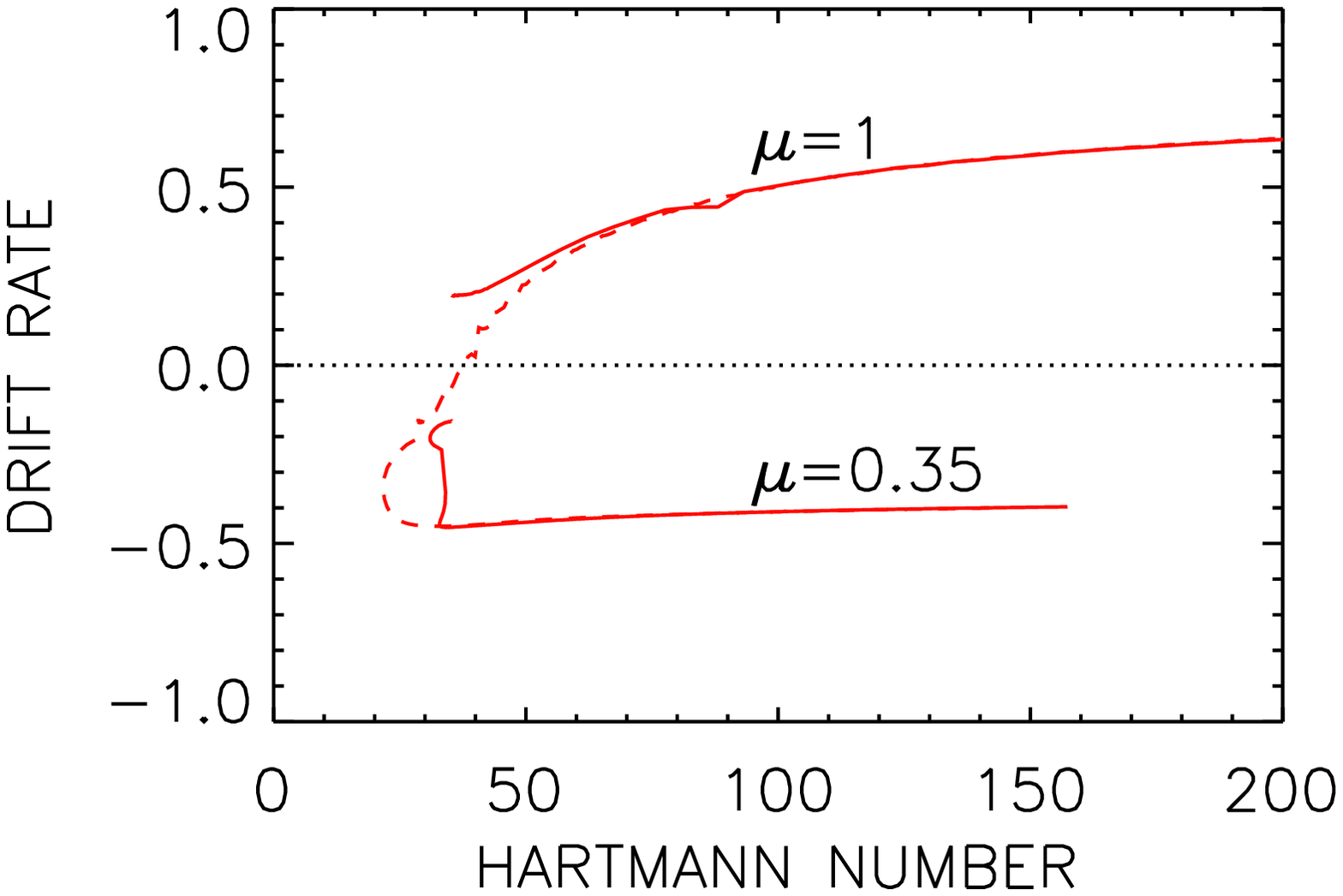}}
   }
  \caption{Left: Stability maps of the  $z$-pinch subject to 
  rigid rotation ($ \mu=1$) and  for
  quasi-Keplerian rotation ($\mu=0.35$). The areas below the lines of neutral stability mark the unstable domains. 
  Middle:  Normalized wave numbers  along the lines of neutral stability. The values  above (below) the horizontal dotted line provide oblate (prolate)  cell structures. Right: Drift rates in units of the inner cylinder rotation along the lines of neutral stability.
  $\rin=0.5$, $\mu_B=2$,  $m= \pm 1$, $\Pm=0.1$. Perfect-conducting boundary  conditions (solid lines),  insulating boundary conditions (dashed lines).  
 }
\label{fig30}
\end{figure*}

On the other hand, the critical Reynolds number is even zero if the toroidal magnetic field is not current-free in the gap between the cylinders and the electric current is strong enough \citep{T57}. The critical amplitudes of the toroidal field form a characteristic Hartmann number  which for resting flows only depends on the radial profile of the field  not  on the magnetic Prandtl number. It takes its minimum if the field is due to an  axial electric current which is homogeneous in the container. Moreover, the Tayler instability is suppressed by the rotation \citep{PT85}. As a consequence, a rotating magnetized Couette flow can only be unstable for supercritical magnetic fields where to all supercritical Hartmann numbers  a maximal Reynolds number belongs above which the instability decays. This line of neutral stability and the instability line for nonrotating flows form a domain of unstable flows as in  the left panel of Fig. \ref{fig30}) which can be probed for their ability to transport angular momentum, to diffuse chemicals,  dissipate  magnetic fields and/or  to form helicity and even $\alpha$ effect  \citep{RG18}.

Flows  of the Chandrasekhar-type,  where the background field and the background flow  have identical  isolines,   are unstable against nonaxisymmetric perturbations if at least one of the diffusivities  is non-zero. For $\Pm\ll 1$ the onset of the instability   also scales with the Reynolds number  and the Hartmann number, i.e. the neutral stability curves converge for $\Pm\to 0$ in the Hartmann number/Reynolds number plane.  A prominent example of this class of magnetohydrodynamic flows is the axially unbounded rigidly rotating $z$-pinch exhibiting  toroidal  flows and fields which only vary with   the distance $R$ from the rotation axis. 
The condition
\beg
\frac{\d}{\d R} (R B_\phi^2) \leq 0
\ende
is  sufficient and necessary  for stability of a stationary ideal fluid against nonaxisymmetric perturbations \citep{T73}. As a consequence, one finds uniform or outwardly increasing magnetic fields unstable against nonaxisymmetric perturbations. This is in particular   true for the field with  $B_\phi\propto R$ produced by   a uniform electric current. The existence of the nonaxisymmetric instability for such a nonrotating  `$z$-pinch' has experimentally been shown by \cite{SS12} using  liquid GaInSn as the conducting fluid penetrated by an axial electric current. 

The Tayler instability in rotating Couette flows  and its   qualification   to induce mean-field electromotive forces shall here  be considered   to question its dynamo activity. It has been  suggested  that the combination of axisymmetric differential rotation and nonaxisymmetric Tayler instability patterns may work as a dynamo  in the convectively stable radiative interiors of stars  \citep{S02}. Numerical simulations by \cite{B06} seem to realize such a dynamo model but sofar its existence  is still under debate \citep{GRE08,ZB07,GT19}. In the present paper we shall attack the problem by use of a mean-field dynamo formulation  considering a  simple $z$-pinch model subject to  differential rotation.  The flows and fields are assumed as unbounded in the axial direction  and the related  instabilities are always nonaxisymmetric. We are here   formulating  a linear theory for the   onset of a possible dynamo and ignore all consequences of its nonlinear evolution. The main result will be that only the assumptions about the numerical values of the Tayler-induced resistivity in its dependencies on magnetic field and  rotation will decide whether a dynamo self-excitation is possible or not. 

By numerical simulations  dynamo excitation   on the basis of the   azimuthal magnetorotational instability (AMRI) of quasi-Keplerian flow has been  shown by \cite{GH17}. Due to the large magnetic Prandtl number $\Pm=10$ the  magnetic energy approaches the kinetic energy which might be a condition of small-scale magnetohydrodynamic dynamo processes. In order to avoid the operation of a small-scale dynamo  mainly $\Pm=0.1$  is used in the present paper while  smaller values would be  beyond the limitations of our numerical calculations. In contrast to  applications with AMRI the azimuthal background field of the Tayler instability is due to a large-scale  axial electric current $\vec J$. A {\em pseudoscalar}  ${\vec B}\cdot {\vec J}$ can thus be formed so that an $\alpha$ effect -- known from the mean-field dynamo theory -- may exist.  Note that the global rotation $\vec \Om$ is not needed for   this argumentation and that for  AMRI with its $\vec J=0$ such a pseudoscalar does not exist at all. 

The outline of the paper is as follows. Section \ref{Equations}  presents  the magnetohydrodynamic equation system which governs the problem. The boundary conditions are given for both insulating and perfect-conducting cylinders.
Section \ref{Pinch} deals with the linear eigenvalue problem of the pinch-type instability where the toroidal field is assumed as due to a homogeneous and axial electric current between the cylindric walls.
The  components of the instability-induced electromotive force  are calculated  in 
Section \ref{Electromotive} including an   $\alpha$ effect  which only appears for a  single-mode but vanishes    if the complete spectrum of the  modes is considered. A mean-field dynamo model is constructed in Section \ref{Dynamo} where by the small value of the resulting helicity the operation of $\alpha^2$ dynamos can immediately be excluded. Section \ref{Discussion} combines the microscopic and the macroscopic results leading to the conclusion that  an $\alpha\Om$ dynamo mechanism is only possible if for the needed fast rotation the instability-induced diffusivity is strongly reduced.

\section{The Equations}\label{Equations}
The  equations of the problem are the standard equations of magnetohydrodynamics, 
\begin{eqnarray}
 \frac{\partial \vec{U}}{\partial t} + (\vec{U}\cdot \nabla)\vec{U}& =& -\frac{1}{\rho} \nabla P + \nu \Delta \vec{U} 
   + \frac{1}{\mu_0\rho}{\textrm{curl}}\vec{B} \times \vec{B},\nonumber\\
 \frac{\partial \vec{B}}{\partial t}&=& {\textrm{curl}} (\vec{U} \times \vec{B}) + \eta \Delta\vec{B}  
   \label{mhd2}
\end{eqnarray}
with $  {\textrm{div}}\ \vec{U} = {\textrm{div}}\ \vec{B} = 0$ for an incompressible magnetized  fluid of density $\rho$.
$\vec{U}$ is the velocity, $\vec{B}$ the magnetic field and  $P$ the pressure. The  basic state in the cylindric system with the
coordinates ($R,\phi,z$) is \mbox{$ U_R=U_z=B_R=B_z=0$} for the poloidal components and 
\beg
\Om  = a  + \frac{b}{R^2}
\label{Om}
\ende
 for the rotation law with  $a$ and $b$ as constants.
$\mur={R_{\rm in}}/{R_{\rm out}}$ be the ratio of the inner cylinder radius $R_{\rm in}$ and the outer cylinder 
radius $R_{\rm out}$ while   $\Om_{\rm in}$ and $\Om_{\rm out}$ are the angular velocities of the inner and outer cylinders, 
respectively. With the definition 
$
\mu={ \Om_{\rm out}}/{\Om_{\rm in}},
$
 sub-rotation (negative shear, ${\rm d} \Om/{\rm d} R<0$) is represented by $\mu<1$ and super-rotation (positive shear, ${\rm d} \Om/{\rm d} R>0$) by $\mu>1$.  The  absolute shear value $|\d \Om/\d R|$  may monotonously sink from the inner cylinder to the outer cylinder. 
 We shall work in this paper  with  uniform rotation ($\mu=1$) as well as with a quasi-Keplerian rotation law ($\mu=\rin^{3/2}$) where the inner and the outer cylinder rotate like planets. 

The two   $z$-independent solutions  of  (\ref{mhd2}) for the magnetic background field $B_\phi$ are $R$ and $1/R$ where the latter is  current-free in the fluid.    We define
$\mu_B=B_{\rm out}/B_{\rm in}$. Only  the  pinch-type solution with $
 B_\phi=B_{\rm in}R/R_{\rm in}$ is here   considered, i.e.  $\mu_B=1/\rin$.

The dimensionless physical parameters of the system besides  the magnetic Prandtl number $\Pm$ are the 
Hartmann number $\Ha$ and the Reynolds number $\Rey$ 
\beg
 {\Ha} =\frac{B_{\rm in} D}{\sqrt{\mu_0\rho\nu\eta}},  \quad\quad\quad
 {\Rey} =\frac{\Om_{\rm in} D^2}{\nu}.
\label{pm}
\ende
where  the inner  magnetic field  and the rotation rate of the inner cylinder are  used.  
The difference   $D=R_{\rm out}-R_{\rm in}$ 
is the gap width between the cylinders. The magnetic Lundquist  number is $\S=\sqrt{\Pm}\cdot \Ha$. 

The variables  $\vec{U}$, $\vec{B}$ and $P$ are split into mean and fluctuating components, i.e. $\vec{U}=\bar{ \vec{U}}+\vec{u}$, $\vec{B}=\bar{ \vec{B}}+\vec{b}$ and $P=\bar P+p$. The bars of the variables are immediately dropped so that the capital letters $\vec{U}$, $\vec{B}$ and $P$ represent the  background quantities. By developing the disturbances $\vec{u}$,  $\vec{b}$ and $p$ into normal modes, 
\beg
[\vec{u},\vec{b},p]=[\vec{u}(R),\vec{b}(R),p(R)] {\rm exp}({{\rm i}(\omega t+kz+ m\phi)}),
\label{fluc}
\ende
the solutions of the linearized MHD equations are constructed for axially unbounded cylinders. Here $k$ is the axial wave number of the perturbation, $m$ its azimuthal wave number and $\omega$ the complex frequency including growth rate as its negative imaginary part and  a drift  frequency $\omega_{\rm dr}$ as its real part. 
A linear code is used to solve the resulting  set of linearized ordinary differential equations 
for the radial functions of flow, field and pressure  fluctuations. The ratio $\eta/D$ serves as the unit of the velocity, $B_{\rm in}$ as the unit of the field perturbations and $\Om_{\rm in}$ as the unit of frequencies. 
The solutions are
optimized with respect to the Reynolds number for given Hartmann number by varying the wave
number. Only solutions for $m=\pm1$ are here discussed. The hydrodynamic boundary
conditions at the cylinder walls are the rigid ones,  i.e. $u_R=u_\phi=u_z=0$. The cylinders are often  assumed to be 
perfectly conducting. For the conducting walls the
fluctuations $\vec{b}$ must fulfill ${\rm d} b_\phi/{\rm d}R + b_\phi/R=b_R=0$ at  $R_{\rm in}$
and $R_{\rm out}$ so that ten boundary conditions exist for the set of ten differential equations.
The magnetic boundary conditions for insulating walls are
\begin{equation}
b_R+\frac{{\rm i}b_z}{I_m(kR)} \left(\frac{m}{kR} I_m(kR)+I_{m+1}(kR)\right)=0
\label{72.7}
\end{equation}
for $R=R_{\rm in}$, and 
\begin{equation}
b_R+ \frac{{\rm i}b_z}{K_m(kR)} \left(\frac{m}{kR} K_m(kR)-K_{m+1}(kR)\right)=0
\label{72.8}
\end{equation}
for $R=R_{\rm out}$, where $I_m$ and $K_m$ are the modified Bessel functions of second kind. The conditions for the toroidal field  are simply $k R b_\phi =m\, b_z$ at $R_{\rm in}$ and $R_{\rm out}$. 
Details including the modified expressions for cylinders with { finite} electric conductivity are given by \cite{RS18}.

\section{Pinch-type instability}\label{Pinch}
The combination of  the  magnetic field $B_\phi\propto R$  and the  rigid-rotation profile $\Om=$~const    belongs  to a particular class of MHD flows defined by  \cite{C56} as
$
 \vec{U}=\vec{U}_{\rm A}.
 $
The radial profiles of flow velocity $\vec{U}$ and  \A\ velocity $\vec{U}_{\rm A}=\vec{B}/ \sqrt{\mu_0\rho}$ are  here identical.  All such flows  are stable in the absence of diffusive effects. 
The rigidly rotating $z$-pinch belongs to the class of the Chandrasekhar-type flows as both  the toroidal  magnetic profile and the linear velocity linearly   run  with  $R$. This implies a uniform axial electric current throughout the entire region $R<R_{\rm out}$.  For all fluids of  the Chandrasekhar-type  in the ($\Ha$/$\Rey$) plane the lines of marginal stability for $m=\pm 1$ converge for  $\Pm\to 0$.

\subsection{Stability maps}
The  curves of neutral instability for $m=\pm 1$  and the two  rotation laws (rigid and quasi-Keplerian rotation) are shown in the left panel of  Fig.~\ref{fig30} for both  perfectly conducting  and insulating  boundaries. 
The resulting instability is purely current-driven. Such instabilities even exist  for $\Rey=0$ for supercritical  Hartmann numbers $\Ha\geq \Ha_0$. The given curves   demonstrate the stabilizing effect of the global rotation   so that for all  $\Ha\geq \Ha_0$  maximal Reynolds numbers $\Rey_{\rm max}$ exist. The curves of the maximal Reynolds numbers become the more steep the smaller the magnetic Prandtl number is. The rotational suppression     is thus strongest for $\Pm=1$, it becomes weaker for smaller magnetic Prandtl numbers.  For  rigid rotation and for quasi-Keplerian rotation it has been shown that for highly supercritical Hartmann numbers  the maximal  Reynolds number grows with growing Hartmann number. For small $\Pm$ and slow rotation also subcritical excitations are possible.

Figure~\ref{fig30} also 
shows the minor importance of the boundary conditions for shape and extension of the instability domains. The rotational stabilization of the nonaxisymmetric instability for differential rotation is much weaker than for rigid rotation.   Above the lines the flow is stable. The limiting  Reynolds number   depends on the magnetic Prandtl number, i.e. the smaller $\Pm$ the higher the maximal  Reynolds number for the instability.  For \C-type flows, however, the $\Rey_{\rm max}$ loose their $\Pm$-dependence if $\Pm\to 0$. This is of course not true for   a quasi-Keplerian flow which allows instability  up to  $\Rey_{\rm max}\simeq 1000$ for $\Pm=1$ and up to $\Rey_{\rm max}\simeq 8812$ for $\Pm=10^{-5}$ (both for   $ \Ha=50$).  Hence, the $\Pm$-dependence of the maximal  Reynolds number is as weak as   $\Remax\propto \Pm^{-0.16}$ for fixed Hartmann number and for small $\Pm$ while this dependence vanishes   for \C-type flows with $\Pm\to 0$.

The middle and right panels of Fig. \ref{fig30} present the normalized wave numbers $kD$ and the normalized drift frequencies  $\omega/\Om_{\rm in}$ along the lines of neutral instability.  The middle panel shows that the shape of the cells  depends on the form of the rotation law. While for rigid rotation the cells are quite oblate they are prolate for the quasi-Keplerian rotation. For stronger magnetic fields the wave numbers monotonously grow. For our two  rotation laws the azimuthal drift rate even possesses opposite signs. Again the influence of the boundary conditions is only weak.

\begin{figure}
  \centerline{
  \includegraphics[width=0.5\textwidth]{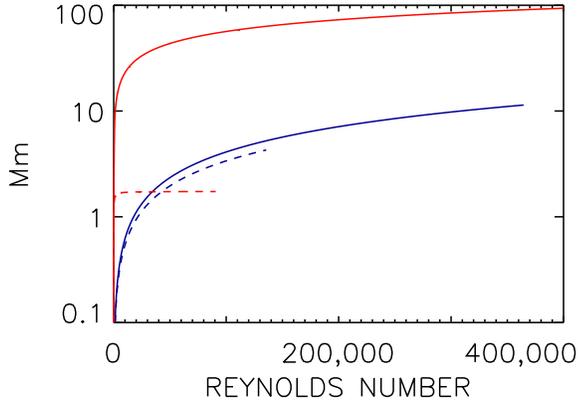}}
  \caption{Magnetic Mach number (\ref{Mm}) versus  Reynolds number $\Remax$ along the lines of marginal stability for quasi-Keplerian rotation and for $\Pm=0.1$ (red line) and $\Pm=10^{-5}$ (blue). The dashed curves are for rigid rotation. Perfect-conducting boundary conditions. 
 }
\label{fig32}
\end{figure}

The shape of the stability lines in the ($\Ha/\Rey$) plane can be expressed by   the  magnetic Mach number $\Mm=\Rm/\Lu$, i.e.
\beg
\Mm=\frac{\sqrt{\Pm}\ \Rey_{\rm max}}{\Ha},
\label{Mm}
\ende 
which for the quasi-Keplerian flow clearly exceeds the value for rigid rotation by one order of magnitude (Fig. \ref{fig32}). For the latter case there is almost no dependence of the curves  on the magnetic Prandtl number.  A weak  dependence  exists  for differential rotation: the magnetic Mach number  varies by one or two orders of magnitude for magnetic Prandtl numbers varying by four orders of magnitude. Only the non-uniform rotation in combination with the  larger $\Pm$ leads to super\A ic values of the $\Remax$, as expected because of enhanced induction in the high-conductivity limit.

The key question for the existence of $\alpha\Om$ dynamos of the Tayler-Spruit-type will be whether the described rotational quenching still allows rotation rates which  are high enough for dynamo excitation driven by the flow and field perturbations.
Also the question arises whether even flows with positive shear ($\mu>1$) may operate as a large-scale dynamo. The Tayler instability under the presence of super-rotation is also suppressed for fast rotation so that the differences to systems with sub-rotation might be small.  For slow rotation   the instability is magnetically  supported so that for $\Pm\neq 1$ subcritical excitations appear \citep{KS12,RG18}. As we shall see, however,  the large-scale dynamo only operates in the fast-rotation limit.
\subsection{Eigensolutions}

The homogeneous system of differential equations for the perturbations forms  an eigenvalue problem with eigensolutions for $\vec{u}(R)$ and $\vec{b}(R)$ which can be determined up to  a free real multiplication factor. 
For $m=\pm 1$ the components $u_R, u_\phi, b_R$ and $b_\phi$ are conjugate-complex as also  the field components $- {\rm i} u_z$ and $- {\rm i} b_z$ are. It means that for the transformation $m\to -m$ the components $b_R$ (and $b_\phi$) transform as $b_R^{\rm R}\to b_R^{\rm R} $ and $b_R^{\rm I}\to -b_R^{\rm I} $  while for  the same transformation   $b_z^{\rm R}\to - b_z^{\rm R} $ and $b_z^{\rm I}\to b_z^{\rm I} $.  The superscripts R stand for the real parts and I for the imaginary parts of the eigensolutions $\vec u$ and $\vec b$. 
These transformation rules  lead to  opposite behaviors of the calculated components of the  electromotive force (such as $\alpha$ effect and eddy resistivity) 
if both modes are excited.

The product of two scalars $A$ and $B$ after averaging over the $\phi$ coordinate is the sum of the products of the real parts and the imaginary parts, i.e. $AB = A^{\rm R} B^{\rm R}+  A^{\rm I} B^{\rm I}$. There is a certain factor in front of this expression whose  value, however,  is unimportant as in the linear theory the  vector components are only known   up to a free factor. It is thus clear that in the linear theory only {\em ratios} of second-order correlations (auto-correlations or cross-correlations) can be calculated which   are free of the arbitrary  parameter.

We shall understand the averaging rules as summation of all values located on a cylinder of radius $R$. The results  are independent of $\phi$ and $z$. This procedure will first be applied to the ratio 
\beg
\varepsilon= \frac{\langle \vec{b}^2 \rangle}{\mu_0\rho\langle \vec{u}^2 \rangle }
\label{eps}
\ende
of the magnetic and kinetic energy. We shall calculate these values as  averages over the whole container along the stability curves $\Rey_{\rm max}=\Rey_{\rm max}(\Ha)$ of the left panel of Fig. \ref{fig30}. In dimensionless quantities it is $\varepsilon= \S^2 \langle \vec{b}^2 \rangle/\langle \vec{u}^2 \rangle $.
\begin{figure}
  \centerline{
  \includegraphics[width=0.5\textwidth]{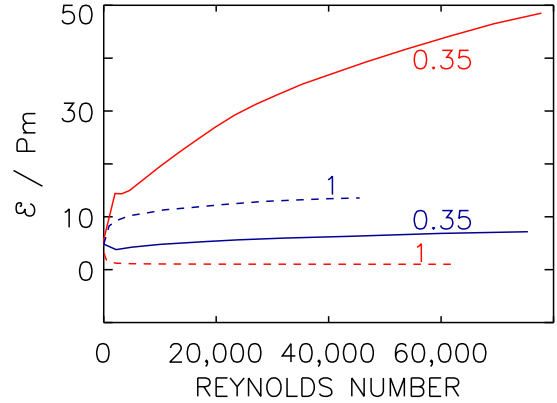}}
  \caption{Ratio $\varepsilon$ after  (\ref{eps}) divided by the magnetic Prandtl number along the line of neutral stability ($\Rey=\Remax$)  for the  pinch with  rigid rotation (dashed lines) and with  quasi-Keplerian rotation (solid lines). The curves are marked with their values of  $\mu$.  $\rin=0.5$, $\mu_B=2$, $m=\pm 1$, $\Pm=0.1$ (red) and $\Pm=10^{-5}$ (blue). Perfect-conducting boundary  conditions. 
 }
\label{fig31}
\end{figure}

Figure \ref{fig31}  gives the results in form of $\varepsilon/\Pm$.  For small $\Pm$ also  the magnetic energy is very 
small compared with the kinetic one, i.e. $b_{\rm rms}\simeq \sqrt{\mu_0\rho\Pm}\ u_{\rm rms}$. The Maxwell stress for $\Pm\ll 1$, therefore,  will not play an important  role in diffusion  processes such as angular momentum transport or turbulent decay of fossil magnetic fields. One finds, however, that for  differential (Kepler)  rotation and larger magnetic Prandtl numbers the magnetic energy indeed  exceeds the kinetic one. The solid red line in the plot (for quasi-Keplerian rotation and $\Pm=0.1$)  clearly demonstrates the generation of magnetic fluctuations by differential rotation   for large $\Pm$, i.e. for large  microscopic electric conductivity.

One also finds that for differential rotation  the $\varepsilon$ slightly  grows with the  Reynolds number  along the line of neutral stability. As expected, the magnetic energy exceeds the kinetic energy the more the faster the rotational shear but this effect is not too strong. 
The energy  ratio $\varepsilon$ in this case  runs as  $\varepsilon\propto \Pm^{0.85}\ {\rm Rm}^{0.35}$. This expression strongly grows with growing electric conductivity indicating the important role of the shear-originated  induction. 

\section{Electromotive force}\label{Electromotive}
Next the axisymmetric part of the  electromotive force ${\vec{\cal E}}=\langle \vec{u}\times \vec{b}\rangle$ is considered which is due to the correlations of flow and field perturbations. We first  proceed again    along the lines of neutral stability  in Fig. \ref{fig30} (left) because a possible dynamo will work much easier for fast rotation than for slow rotation.
The line of neutral stability defines the maximally possible rotation rates  belonging to a given magnetic field. 
Below this line the  rotation is  slower so that for dynamo excitation the $\alpha$ effect must be larger. 

By proper normalizations the influence of the  unknown free factor can be eliminated. 
Again the averaging procedure concerns the time and the  coordinates $\phi$ and $z$. The mean-field  electromotive force may be developed in the  series
\beg
{\vec{\cal E}}= \alpha \vec{B}- \eta_{\rm T} \curl \vec{B}+....
\label{EMF}
\ende
with the $\alpha$ effect and the eddy diffusivity   $ \eta_{\rm T}$ as coefficients of the large-scale  field and the large-scale electric current which both are  nontrivial tensors. In cylinder geometry, the $\phi$-component
$
{\cal E}_\phi= \langle u_z b_R -  u_R b_z\rangle 
$
 can be written as  the axisymmetric part of expressions  such as $\langle u_z b_R\rangle\propto u_z^{\rm R} b_R^{\rm R}+u_z^{\rm I} b_R^{\rm I}$. Because of the different transformation rules for radial and axial components of the eigenvectors the expression ${\cal E}_\phi$ takes opposite signs for $m\to -m$. 
It is thus evident that the total azimuthal  electromotive force due to the   instability of azimuthal fields  vanishes if both modes $m$ and $-m$ (which have the same eigenvalues and the same azimuthal drifts) are simultaneously excited with the same power. Only   by an extra parity braking (e.g. by an additional $z$-component of the magnetic background field) finite values of the $\alpha$ effect appear. Another possibility is to consider the isolated modes $m=1$ and $m=-1$   as the result of  a spontaneous parity braking or  by use of strictly formulated initial conditions. 
 If the initial conditions clearly  favor one mode then only   this one is excited. If the initial condition do not favor one of the two modes, the numerical noise will determine the dominant mode.  Not necessarily the 
solutions  consist of equal mixtures of both degenerated modes \citep{GR11,CM11,BB12}.
In this case the isolated mode with $m=1$ possesses a finite helicity (even without rotation) which is the negative value of the helicity of the mode $m=-1$.
 \begin{figure}
  \centerline{
 \vbox{  \includegraphics[width=0.45\textwidth]{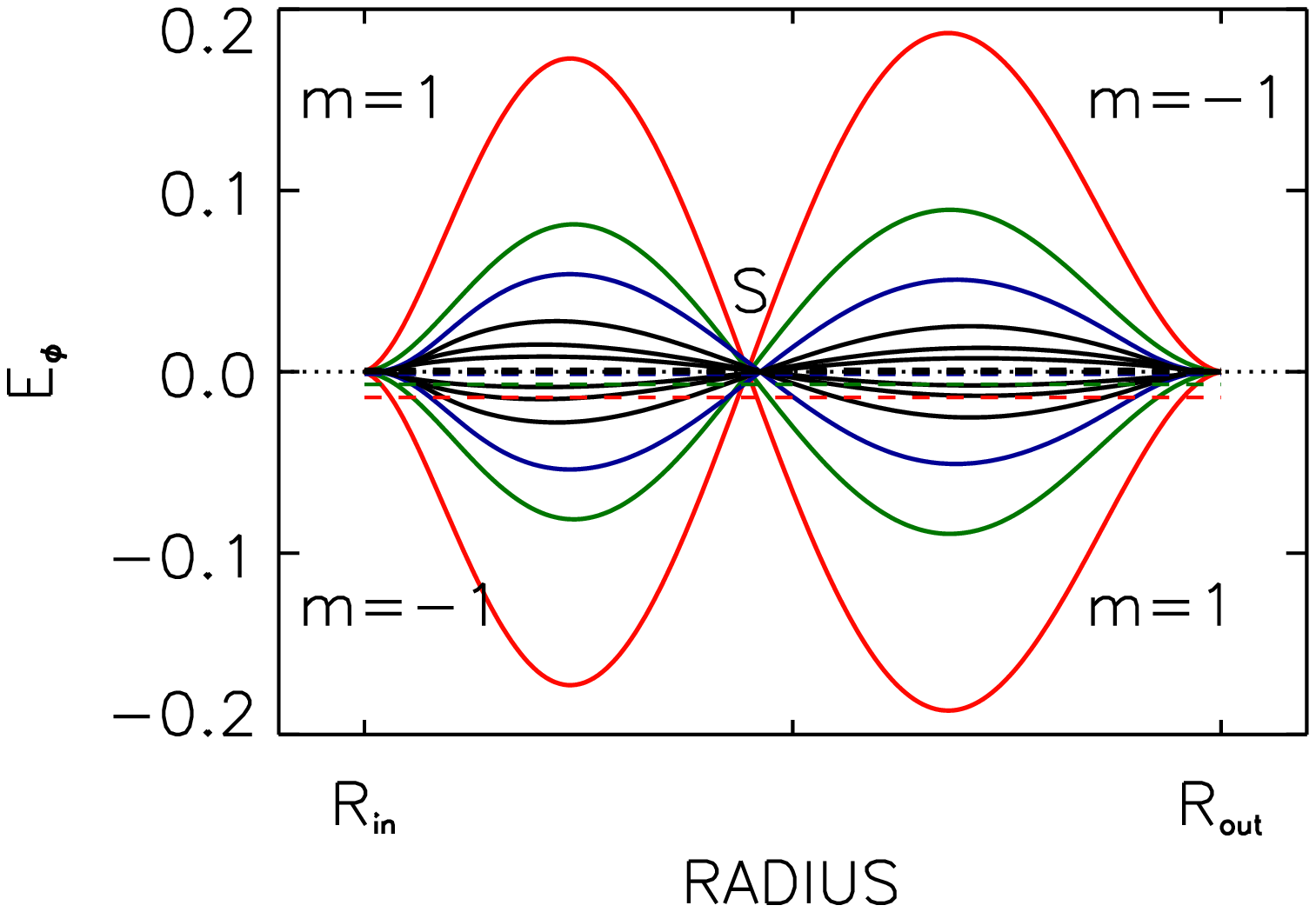}
  \includegraphics[width=0.45\textwidth]{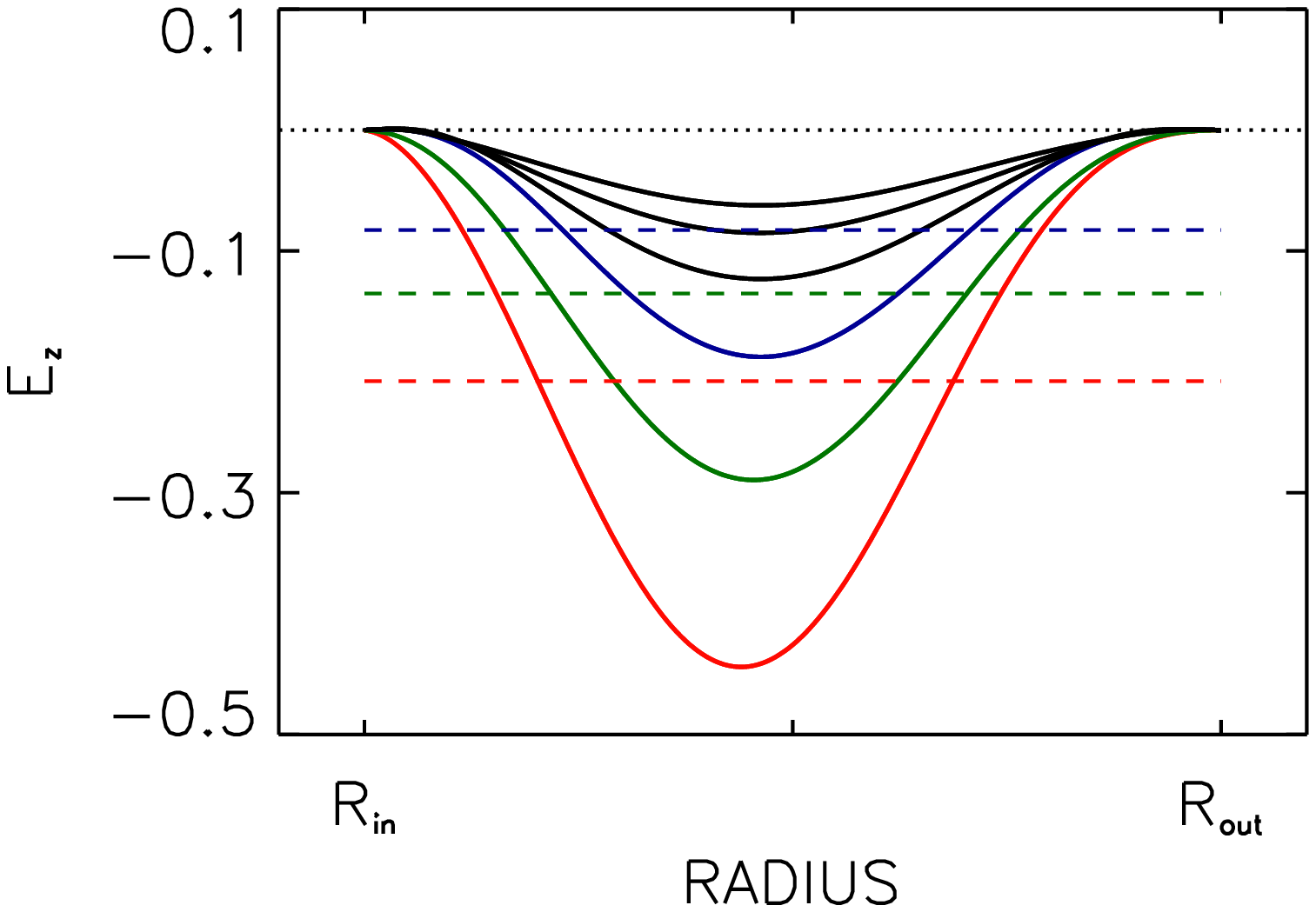} }}
   \caption{Normalized electromotive force ${\cal E}_\phi$ (top) and ${\cal E}_z$ (bottom) along the line of neutral stability  within   rigidly-rotating   $z$-pinches.  It is $\Rey=0$, $\Ha=35$ (no rotation, red),  $\Rey_{\rm max}=140$, $\Ha=50$ (slow rigid rotation, green) and $\Rey_{\rm max}=418$, $\Ha=100$ (fast rigid rotation, blue). The horizontal dashed lines give  the values averaged over the radius for $m=1$. It is $\cal E_\phi\to -\cal E_\phi$ and ${\cal E}_z\to {\cal E}_z$ for $m\to -m$. $\rin=0.5$, $\mu_B=1/\rin=2$,  $m=\pm 1$.  Perfect-conducting cylinder walls. --   We shall call the  profiles of the top panel  for $m=1$ as the ``sine-type'' $\alpha$ effect  and for $m=-1$ as the ``minus-sine-type''. S denotes the  zeros of the sine-profiles which do not depend on $m$ and $\Rey$.  $\Pm=0.1$.}
\label{fig34a}
\end{figure}

The $z$-component of the mean-field  electromotive force  in cylinder geometry is
$
{\cal E}_z= \langle u_R b_\phi -  u_\phi b_R\rangle.
$
The radial and azimuthal components of flow and field  are  invariant against the transformation $m\to -m$. Both the azimuthal and axial  components of  ${\vec{\cal E}}$ for resting or rigidly  rotating containers  are given in Fig. \ref{fig34a} normalized with the maximal  total energy MAX$(\langle \vec{u}^2 + \vec{b}^2/\mu_0\rho\rangle)$. 
For fixed $m$  all curves for  ${\cal E}_\phi$  are {\em almost} antisymmetric with respect to the common zero marked by S. This is a direct consequence of the  one-cell pattern of the instability in  linear approximation. Always the radial components of $\vec u$ and $\vec b$ are symmetric with respect to S and the axial components are antisymmetric resulting in an antisymmetric structure of  ${\cal E}_\phi$. 
For fixed $m$ the  ${\cal E}_\phi$ curve behaves antisymmetric with respect to $m\to -m$ but  the  ${\cal E}_z$ curve does not.  Here the curves for $m=1$ and $m=-1$ are identical and hence the eddy diffusivity $\eta_{\rm T}$ is a  robust quantity. The negative-definite sign of ${\cal E}_z$ will lead to the expected positive-definite  sign of $\eta_{\rm T}$.

\subsection{Alpha effect}\label{Alpha}
Consider   the components ${\cal E}_\phi$ and   ${\cal E}_z$  of the electromotive force. The ratio of both quantities is free of arbitrary factors and/or  normalizations. 
We write $\varepsilon_\alpha={\cal E}_\phi/{\cal E}_z$, i.e.
\beg
\varepsilon_\alpha=\frac{ \langle u_z b_R -  u_R b_z\rangle }{\langle u_R b_\phi -  u_\phi b_R\rangle}.
\ende
With ${\cal E}_\phi=\alpha B_\phi$ and ${\cal E}_z=-\eta_{\rm T} \curl_z \vec{ B}$ one immediately finds 
$
\varepsilon_\alpha= {C^{\rm sim}_\alpha}/{2},
$
where the standard notation
\beg
C^{\rm sim}_\alpha= \frac{\alpha \Rin}{\eta_{\rm T}}
\label{EMF3}
\ende
has been used.  For uniform $\alpha$ effect the excitation of a large-scale dynamo  in an axially unbounded resting cylinder    needs $C_\alpha>1$   \citep{M90}. 

The radial profiles of the two  $\vec{\cal E}$~components are given in Fig. \ref{fig34a} for the stationary pinch (red lines) and the rigidly rotating pinch (green and blue lines)  for marginal stability, for $\Pm=0.1$ and  for the   modes with $m=\pm 1$.   Averaged  over $\phi$ and $z$ (defining cylindrical surfaces) the ${\cal E}_\phi$  is sinusoidal.
S denotes the nulls of the quasi-sine-profiles which are shown as independent of  $m$ and  $\Rey$. One  also finds  that the {\em radial} average of the  ${\cal E}_\phi$ 
does not vanish as the amplitudes of the  outer parts of the curves slightly exceed those of  the inner parts.  Averaged over the entire container, therefore,  the ${\cal E}_\phi$ is uniform but small   with   opposite signs for $m=1$ and $m=-1$. 
 It is $C_\alpha=O(0.1)$  if averaged over the whole container. 
The blue lines in Fig. \ref{fig34a} hold for stronger magnetic fields with $\Ha=100$ and  Reynolds number $\Remax=418$. The dashed lines represent the corresponding values as averaged over the entire container.
Note that the influence of the global rotation is only weak. The dependence of $C_\alpha$ on the radial distribution $\mu_B$ is not yet  known.

\begin{figure}
  \centerline{
 \vbox{
  \includegraphics[width=0.50\textwidth]{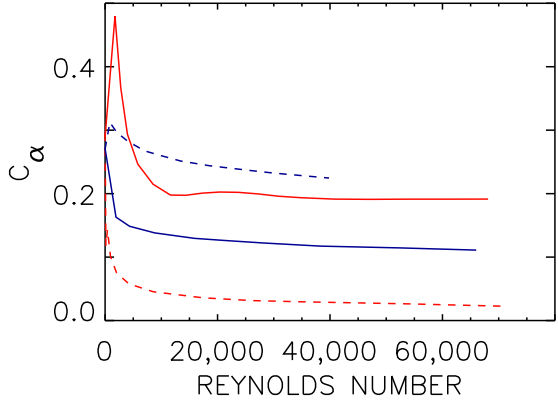} }
  }
  \caption{Simulation results of  $C ^{\rm sim}_\alpha$ averaged over the entire container for rigid rotation ($\mu=1$, dashed lines) and quasi-Keplerian rotation ($\mu=0.35$, solid lines). $\Pm=10^{-5}$ (blue) and $\Pm=0.1$ (red). $\rin=0.5$, $\mu_B=2$,  $m= -1$.  Perfect-conducting cylinder walls.
  }
\label{fig35}
\end{figure}

The ratio $\varepsilon_\alpha$ identically  vanishes if both spirals with $m=1$ and $m=-1$ are excited with the same power. A finite $\alpha$ effect  exists if only one of these modes is excited.
Figure \ref{fig35}
shows the simulated $C^{\rm sim}_\alpha $   along the lines of neutral stability  for  uniform rotation ($\mu=1$, dashed lines) and quasi-Keplerian rotation ($\mu=0.35$, solid lines) as  functions of the maximal  Reynolds number. The magnetic Prandtl number varies by four orders of magnitude between $\Pm=0.1$  and $\Pm=10^{-5}$. 
For not too slow rotation the resulting  $C^{\rm sim}_\alpha $  hardly depend on the magnetic Prandtl number,  the rotation law {\em and} the Reynolds number.  For Kepler rotation   $C^{\rm sim}_\alpha\simeq 0.1...0.2$ is the characteristic result of the simulations. 
 The rotational suppression proves to be   surprisingly  weak. The same holds for  the $\Pm$-dependency.
The red lines in Fig. \ref{fig35} also show that for $\Pm=0.1$ the differential rotation produces $C^{\rm sim}_\alpha$ larger by a factor of two than  for uniform rotation. For small $\Pm$ these differences are even reduced.

We find that  without  and with  rotation the pinch-type instability possesses zero $\alpha$ effect if all modes are taken into account but it possesses $C^{\rm sim}_\alpha$ values of $O(0.1)$  if only one of the  modes  is considered. The simulated $C_\alpha$ are not quenched by the rotation but the resting $z$-pinch possesses slightly larger normalized $\alpha$ effect than the rotating one.

\begin{figure}
  \centerline{
 \includegraphics[width=0.50\textwidth]{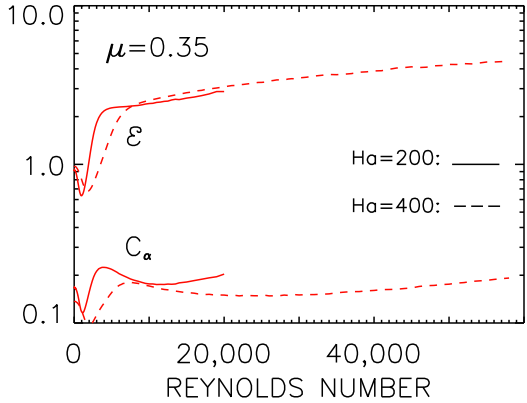} }
  \caption{The normalized $\alpha$ effect $C_\alpha^{\rm sim}$ and the energy ratio $\varepsilon$ along a vertical line in the stability map (Fig. \ref{fig30})  for   two  given Hartmann numbers. It is always  $0\leq \Rey\leq  \Rey_{\rm max}$. Quasi-Keplerian rotation, $\mu=0.35$. $\Pm=0.1$, perfect-conducting walls.
  }
\label{figvertical}
\end{figure}

It remains to ask whether the dynamo excitation is more easy for fixed Hartmann number but for slower rotation, i.e. for  $\Rey<\Rey_{\rm max}$. The calculations along vertical lines in the stability map of  Fig. \ref{fig30} are much more complicated than they are along the line of neutral instability. For $0\leq \Rey<\Rey_{\rm max}$ (at given $\Ha$) the growth rates of the magnetic  instability  are finite. The wave numbers and the drift frequencies must be  optimized to find  the maximal growth rates. The eigensolutions have been computed for exactly these values.  We note that the absolute maximum of the  growth rate exists for very small Reynolds numbers where $\alpha\Om$ dynamos do certainly not exist. Also   vertical slices through the instability domain  do not provide strong dependencies of the simulated $C_\alpha$ on the Reynolds number (Fig. \ref{figvertical}). The solid lines in this graph belong to $\Ha=200$ while the dashed lines belong to $\Ha=400$. The result is that from a minimal  Reynolds number on,   the entire instability domain of Fig. \ref{fig30}  possesses more or less the same value   of $C_\alpha^{\rm sim}=O(0.1)$. 
Any mean-field dynamo theory might base on this basic result. 

Also the energy ratio $\varepsilon$ along the two vertical slices has been  given in Fig. \ref{figvertical}. The results are similar to those  for the $\alpha$ effect but for resting pinches  there is a clear minimum with equipartition of the two energies. For increasing rotation the  $\varepsilon$  increases up to the values  with    $b_{\rm rms}> \sqrt{\mu_0\rho}\ u_{\rm rms}$   which is   already known  for $\Rey=\Remax$ (Fig. \ref{fig31}). Obviously, the numbers derived for the line of neutral instability well represent the numbers valid for the entire instability domain.
 \subsection{Eddy diffusivity}\label{Eddy}
 We shall turn now to the axial component of the electromotive force normalized with the kinetic and magnetic energies, i.e.
 \beg
\varepsilon_z= \frac{\langle u_R b_\phi-u_\phi b_R\rangle}{\sqrt{\langle\vec{u}^2\rangle\langle\vec{b}^2\rangle}}.
\label{diff1}
\ende
Nominator and denominator are of the same dimension. This ratio  can be computed with the quasilinear code in the same way as described above. It is always negative and does hardly depend on the rotation law.  The dependence on  the magnetic Prandtl number is $\varepsilon_z\propto \Pm^{-1/2}$ (except for $\Rey=0$).

\begin{figure}
  \centerline{
  \includegraphics[width=0.50\textwidth]{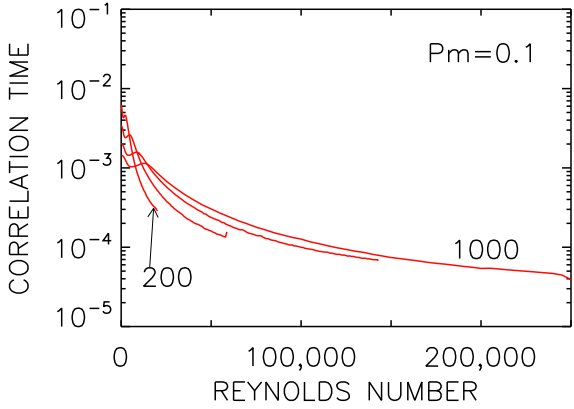} }
  \caption{Simulation results of the correlation time  $\hat\tau_{\rm corr}$ 
  for  quasi-Keplerian rotation. Four fixed Hartmann numbers $\Ha=200$, $\Ha=400$, $\Ha=700$, $\Ha=1000$, as marked. $\mu=0.35$,  $\Pm=0.1$. 
   Perfect-conducting cylinder walls. }
\label{fig36a}
\end{figure}

Replacing the nominator in Eq. (\ref{diff1}) by means of the diffusion approximation one finds with (\ref{EMF})
\beg
\etaT= -\frac{\varepsilon_z}{2}\sqrt{\mu_0\rho\varepsilon}\langle\vec{u}^2\rangle\frac{\Rin}{B_{\rm in}}.
\label{diff2}
\ende
Transformed to code units via $\vec{u}=\hat{\vec{u}}\ \eta/D$ to magnetic Reynolds numbers one obtains
\beg
\frac{\etaT}{\eta}= \frac{\sqrt{\varepsilon^2_z\varepsilon}}{2\S} {\hat u}_{\rm rms}^2
\label{diff2a}
\ende
with ${\hat u}_{\rm rms}=\sqrt{\langle{\hat{\vec{u}}}^2\rangle}$ as the turbulence velocity in code units. 
The magnetic energy of the perturbations does not explicitely contribute to the eddy diffusivity \citep{VK83}. With the standard approximation $\etaT\simeq \tau_{\rm corr} u^2_{\rm rms}$ for an eddy diffusivity one  finds  the correlation time $\tau_{\rm corr}$ as
\beg
\tau_{\rm corr}\simeq  \frac{\sqrt{\varepsilon_z^2\varepsilon}}{2} \frac{D}{V_A},
\label{diff3}
\ende
which linearly scales  with the magnetic turnover time  $D/V_A$ ($V_A$ the  \A\ velocity of the toroidal field). The factor in front of $D/V_A$ in (\ref{diff3})   with 0.01...0.1 is  only small with a very weak  variation  with the magnetic Prandtl number. The factor at the r.h.s. side of Eq. (\ref{diff2a})  can also be understood as a normalized correlation time,
\beg
{\hat\tau}_{\rm corr}\simeq  \frac{\sqrt{\varepsilon_z^2\varepsilon}}{2\Lu}.
\label{diff2aa}
\ende
The ratios $\varepsilon$ and  $\varepsilon_z$ have numerically been  calculated  for several fixed Hartmann numbers in its dependence on the rotation rate and then  combined to $\hat\tau_{\rm corr}$ (Fig. \ref{fig36a}). The curves end at  $\Rey=\Remax$ where $\hat\tau_{\rm corr} \Lu\simeq 0.02$. The correlation time decreases for faster rotation, it  varies by one or  two orders of magnitudes. The results indicate that   the eddy diffusivity for given Hartmann number  close to the line with $\Rey=\Remax$ is basically smaller than for slow rotation. Without rotation one finds ${\hat \tau}_{\rm corr} \Lu \simeq 0.3$ independent of the  magnetic Prandtl number.

 Only  a  nonlinear 3D MHD code can provide  the turbulence intensity $\langle\vec{u}^2\rangle$ in its dependencies on the rotation and the  magnetic field. Below we shall see that  just this  dependence   decides { whether the instability can  work as a dynamo} or not.  
In a first approximation, the turbulence intensity  may be estimated with the following argument. Let us only consider -- as is often assumed in dynamo theory --  those containers where the turbulence is in equilibrium with the applied magnetic field, i.e.
$
\mu_0\rho u_{\rm rms}^2= \kappa_ {\rm eq} B_{\rm in}^2,
$
with a dilution factor $\kappa_ {\rm eq}\leq1$, hence
\beg
\frac{\etaT}{\eta}= \hat\tau_{\rm corr}\ \kappa_ {\rm eq}\ {\rm S}^2,
\label{diff4}
\ende   
which  leads to a {\em linear dependence} of the eddy diffusivity on the magnetic field. 
Figure \ref{fig36a} also demonstrates a clear rotational decay of the correlation time.  We note that numerical simulations of the magnetic diffusivity for quasi-Keplerian rotation and for a  variety  of magnetic Prandtl numbers indeed for large $\Lu$ lead to  $\etaT/\eta\lsim \Lu/10$  \citep{RG18}.


A more detailed  turbulence model results from direct numerical simulations of the pinch-type instability with a nonlinear MHD code.  The turbulence intensity $u^2_{\rm rms}$ depends on the magnetic field  and the rotation rate. Here we shall  only use  the $\Lu$ dependence while the rotation dependence   is still ignored. 
One finds a steep  growth of the turbulence intensity with $\Lu$, i.e.
\beg
\hat u_{\rm rms}^2= \kappa  \Lu^4.
\label{diff6}
\ende   
To date the relation (\ref{diff6}) is only known  for the lowest possible Lundquist numbers.
One finds 
  \beg
\frac{\etaT}{\eta}= \hat\tau_{\rm corr}\ \kappa  \Lu^4.
\label{diff7}
\ende   
The $\kappa$  defines that Lundquist number $\S_1$ where $\etaT/\eta=1$. Preliminary nonlinear simulations by \cite{RG18}  led to $\S_1\simeq 100$  so that from (\ref{diff7}) $\kappa\simeq 5\cdot 10^{-5}$ results. 
The specific eddy diffusivity  exceeds unity if  $\S> \S_1$. There the ratio of $u_{\rm rms}$ to the linear rotation velocity of the cylinder is a few percent. Only for even stronger magnetic fields   $\etaT/\eta\gg 1$ is possible.

In this model  the eddy diffusivity  grows rapidly  as $\Lu^3$ with the magnetic field amplitude.  Also the  Tayler-Spruit dynamo model in its original formulation works with an eddy diffusivity growing cubic  with the magnetic background field and decaying with $\Om^{-1}$ \citep{S02}. 


\section{Dynamo equations}\label{Dynamo}
The possibility will now  be checked whether  the perturbation patterns of the Tayler instability in combination with differential rotation may work as a dynamo  leading to  an axisymmetric  large-scale magnetic field  with  dominating azimuthal component. The model of the rotating $z$-pinch may be  the most simple one with which this problem  can be attacked.
To consider dynamo action for more complicated models is certainly more complicated. 

The mean-field dynamo equation is
\beg
 \frac{\partial \vec{B}}{\partial t}= {\textrm{curl}} (\alpha\cdot \vec{B}+\vec{U} \times \vec{B}) + (\eta+\eta_{\rm T}) \Delta\vec{B} . 
   \label{dyn1}
\ende
The quantities $\alpha$ and $\etaT$  are derived by an averaging over time and space, $\eta+\etaT$ is assumed as uniform. As in contrast  to convection the eddy resistivity  $\etaT$ by magnetic instabilities is not necessarily large compared to $\eta$ we have left the molecular resistivity  in the mean-field equation  (\ref{dyn1}). 
It  shall be solved with the divergence-free ansatz $\vec{B}=\curl (A(R,z) \vec{e}_\phi) + B(R,z)\vec{e}_\phi $ for axisymmetric large-scale fields. The vector $\vec{e}_\phi$ is the unit vector in azimuthal direction. The temporal and also the   $z$-dependence  for both components $A$ and $B$ are considered in the Fourier space, i.e. they are assumed as proportionate to $\exp{\i}(\omega t+ Kz)$. The real part of $\omega$ is the cycle  frequency of the dynamo. In order to have established a mean-field dynamo the resulting wave number $K$ must be  smaller than $1/D$. 

The special case $K=0$ describes a dynamo without any $z$ dependence.
Then the  radial field  $B_R=- {\rm i} K A$ identically  vanishes. In this case the differential rotation cannot induce new field components and the $\Om$ term in the dynamo equation vanishes. For $K=0$, therefore,  only $\alpha^2$ dynamos can exist. The question is thus whether axially-periodic dynamos operate for finite values of the  wave numbers $K$ so that  the $\alpha^2$ dynamo type becomes an $\alpha\Om$ dynamo type which is even able to operate (for fast rotation) with  rather weak $\alpha$ effect.
 The equations for $A(R)$ and $B(R)$ are
\beg
{{\rm d}^2 A \over {\rm d} R^2} 
+ \frac{1}{R}
{{\rm d} A \over{\rm d} R}
 -{A \over R^2}  -({\i}\omega+K^2)  A 
 + C_\alpha {\hat \alpha} B = 0
\label{dyn2}
\ende
for the poloidal field component and 
 \beg
 \begin{split}
{{\rm d}^2 B \over {\rm d} R^2} 
+ {1\over R}
{{\rm d} B \over{\rm d} R}
 -{B \over R^2}  -({\i}\omega+K^2)  B + &  \hat\alpha C_\alpha K^2 A - \\\
 -C_\alpha {\d \over \d R}\left({\hat\alpha}({\d A\over \d R}+{A\over R})\right)
 - & \i K C_\Om  R {{\rm d} {\hat \Om} \over {\rm d} R} A = 0
 \end{split}
\label{dyn3}
\ende
for the toroidal component. Length scales have been normalized by the gap width  $D$, frequencies with the diffusion frequency $(\eta_{\rm T}+\eta) /D^2$ and  the normalized  wave number is $D K$. The profiles $\hat \alpha$ and $\hat \Om$ are normalized 
to unity. $C_\alpha$  has been defined with (\ref{EMF3}) as $ C_\alpha=C_\alpha^{\rm sim} /(1+(\eta/\eta_{\rm T}))$ and $C_\Om$ is the standard magnetic Reynolds number of rotation, $C_\Om=\Om_{\rm in} D^2/(\eta+\eta_{\rm T})$.
One can thus write 
\beg
C^{\rm sim}_\alpha= \left(1+\frac{\eta}{\eta_{\rm T}}\right) C_\alpha,  \ \ \ \ \ \ \ \ \  \Rm=\left(1+\frac{\eta_{\rm T}}{\eta}\right) C_\Om,
\label{sim}
\ende 
hence 
$
C_\alpha<C^{\rm sim}_\alpha$ and    $ \Rm>C_\Om$.
For $\alpha\Om$ dynamos with $C_\Om\gg C_\alpha$ only the product 
\beg
{\cal D}=C_\alpha C_\Om
\label{D} 
\ende
forms the relevant eigenvalue indicating that dynamos are possible even for  small $\alpha$  if only the rotation is  fast enough.
\subsection{Boundary conditions}
For perfectly conducting cylinders the boundary conditions are
 \beg
 A=0,\ \ \ \ \ \ \ \ \ \ \ \ \ \  \frac{\d B}{\d R}+\frac{ B}{R}-C_\alpha {\hat \alpha}\left(\frac{\d A}{\d R}+ \frac{A}{R}\right)=0 
 \label{pc}
 \ende
  at $R=\Rin$ and $R=\Rout$. The first condition  ensures the vanishing of the radial magnetic field  inside the inner cylinder and outside the outer cylinder while the second condition produces zero tangential component of the electromotive force. 
Pseudovacuum conditions (also called vertical field conditions) would require  
$\d A/\d R+ A/R=B=0$ at $R=\Rin$ and $R=\Rout$ \citep{J14}. 
For  more  heuristic dynamo  models also the simplified  conditions $B_R=B_\phi=0$,  i.e. $A=B=0$, have been  used (Roberts 1972).

We shall also  solve the dynamo equations (\ref{dyn2}) and (\ref{dyn3}) with the insulating boundary conditions, i.e. with   (\ref{72.7}) and (\ref{72.8}) taken for $m=0$, i.e. 
 \begin{equation}
 \left(\frac{I_{1}(KR)}{I_0(KR)}-K R\right)A +\frac{I_1(K R)}{I_0(KR)} R\frac{\d A}{\d R}=0,\ \ \ \ \ \ \  B=0
\label{BC1}
\end{equation}
for $R=R_{\rm in}$, and 
\begin{equation}
\left(\frac{K_{1}(K R)}{K_0(K R)}+K R\right)A +\frac{K_1(K R)}{K_0(K R)} R\frac{\d A}{\d R}=0,\ \ \ \ \ \ \  B=0
\label{BC2}
\end{equation}
for $R=R_{\rm out}$. Again  $I_m$ and $K_m$ are the modified Bessel functions of second kind. 
\subsection{An analytical model}
For uniform $\alpha$  effect ($\hat\alpha=1$) in a resting container  and for $\omega=0$  a stationary  analytic solution     $B=C_\alpha A$ of Eqs. (\ref{dyn2}) and (\ref{dyn3}) exists  if 
\beg
{{\rm d}^2 A \over {\rm d} R^2} 
+ {1\over R}
{{\rm d} A \over{\rm d} R}
 + ({\tilde C}_\alpha^2-{1\over R^2}) A = 0
\label{dyn4}
\ende
possesses an  eigenvalue where  ${\tilde C}_\alpha^2=C_\alpha^2-K^2>0$ \citep{M90}. The boundary condition $A=0$ at $\Rin$ and $\Rout$ from Eq. (\ref{pc}) are used  (which also means $B=0$) producing   the eigenequation
\beg
J_1(\tilde C_\alpha)Y_1(\tilde C_\alpha/\rin)- J_1(\tilde C_\alpha/\rin)Y_1(\tilde C_\alpha)=0,
\label{dyn5}
\ende
 where $J_1$ and  $Y_1$ are  the Bessel  functions  of first kind with index $m=1$.
The  $Y_n$ are also called the Neumann-Weber functions. For $\rin=0.5$ one obtains from (\ref{dyn5}) the approximative eigenvalue $\tilde C_\alpha\approx \pm \pi$. The  numerically derived  exact value   is $\tilde C_\alpha=\pm 3.17$ hence $C_\alpha =\sqrt{10.05+K^2}$ so that $C_\alpha$ grows with growing wave number. The axially  homogeneous dynamo without axial bounds 
 is  obviously easiest to excite. The corresponding  kinematic dynamo field is stationary,  axisymmetric and helical.
This analytical  solution, however, is  only of academic interest as  the  simulated $\alpha$ effect averaged over the container   is much smaller than  $C_\alpha=O(\pi)$ required for the cylindrical $\alpha^2$ dynamo.
\subsection{Numerical solutions}
\begin{figure}
\centering
\vbox{
 \includegraphics[width=0.45\textwidth]{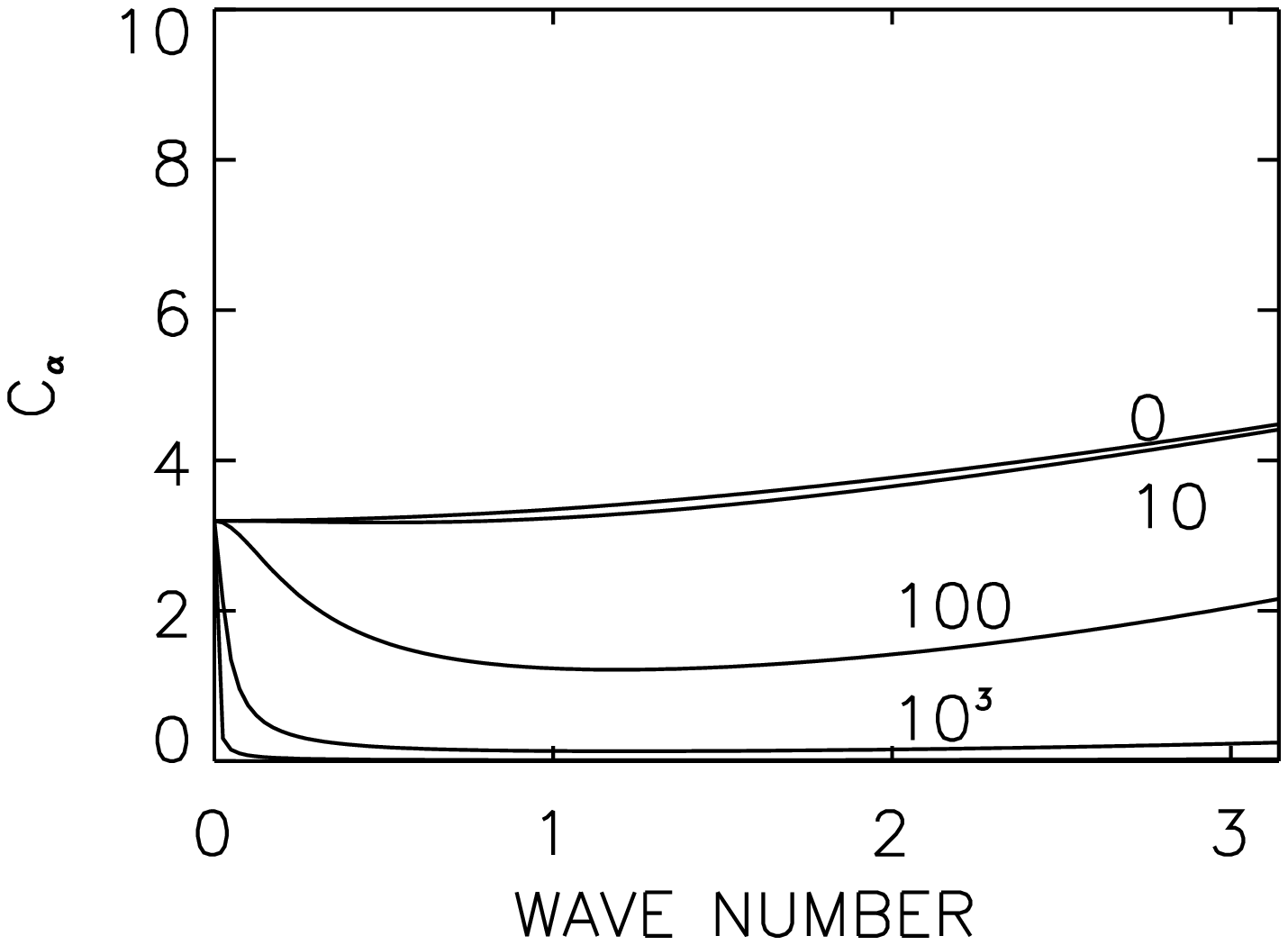}
 \includegraphics[width=0.45\textwidth]{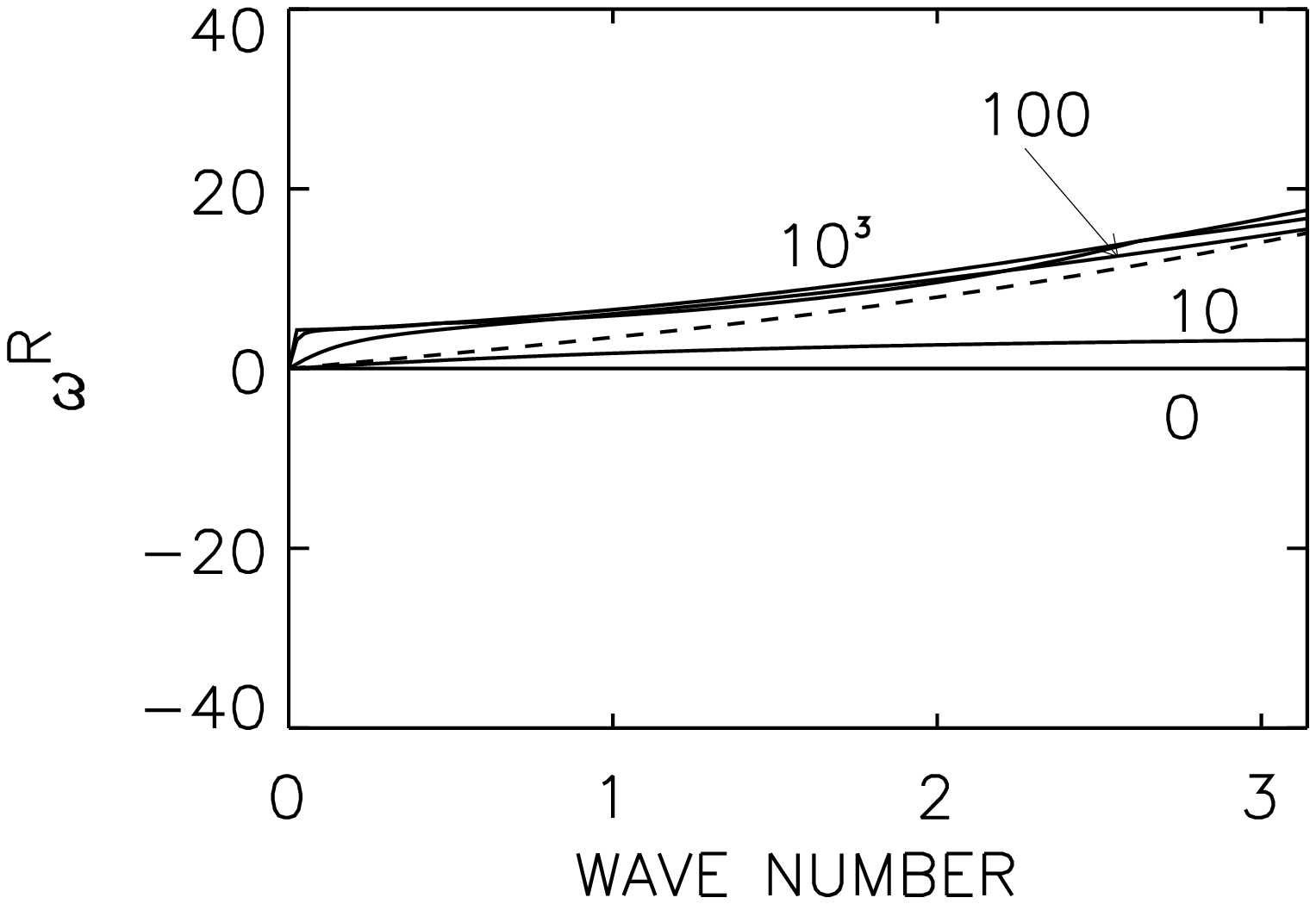}
 }
  \caption{Eigenvalues $C_\alpha$ (top) and drift frequencies (bottom)  of $\alpha^2\Om$ dynamos with  quasi-Keplerian rotation for insulating walls
  and  uniform $\alpha$ profile.
  The curves are marked with their values of  $C_\Om$. $\rin=0.5$,  $\mu=0.35$. }
\label{fig37}
\end{figure}

The numerical  solutions of the  equations (\ref{dyn2}) and (\ref{dyn3}) for infinite cylinders with finite $C_\Om$ have been obtained for three different radial $\alpha$ profiles. Quasi-Keplerian rotation with $\mu=0.35$ for $\rin=0.5$ is always used. The upper panel of  Fig. \ref{fig37} gives for insulating boundary conditions the eigenvalues $C_\alpha$ as function of the wave number and the normalized rotation rate $C_\Om$. We note the choice $\alpha=-1$.  Reynolds number $C_\Om$ of rotation  and  wave number $K$ are the   free model parameters. The curves are marked with their value of $C_\Om$. For $C_\Om=0$ the analytical solution $C_\alpha =\sqrt{10.05+K^2}$ is well approximated where $K=0$ produces the minimal $C_\alpha$. 
For slow rotation ($C_\Om\lsim 10$) the influence of the differential rotation is only weak  and the $C_\alpha$ slightly grows with growing wave number. For larger $C_\Om$ the eigenvalue $C_\alpha$ sinks with growing wave number and possesses   minima moving to smaller $K$ for faster rotation. 
For $C_\Om=1000$ the minimum value is $C_\alpha\simeq 0.13$.  The dynamo numbers taken at the minimum thus converge to ${\cal D}=130$. 

The lower panel of Fig. \ref{fig37} presents the cycle frequency $\omega^{\rm R}$ of the dynamo normalized with the magnetic-dissipation frequency $(\etaT+\eta)/D^2$. The sign of this quantity depends on the sign of the $\alpha$ effect, it is positive for negative $\alpha$. For vanishing $K$ the dynamo turns into an $\alpha^2$ dynamo which does not oscillate.
It oscillates, however, already for slow differential rotation with $C_\Om=10$. For faster rotation the cycle frequency slightly  grows. The frequency always exceeds  the dissipation frequency but in all cases the ratio $\omega^{\rm R}/C_\Om$ (which gives the dynamo  frequency in units of the rotation rate) is small.   Drift rates and growth rates of the magnetic instabilities always scale with the rotation rate. If the dynamo exists then the drift of the instability pattern of Tayler instability is short compared with the cycle frequency, at least for $C_\Om\gsim 100$.

\begin{figure}
\centering
\vbox{ 
\includegraphics[width=0.45\textwidth]{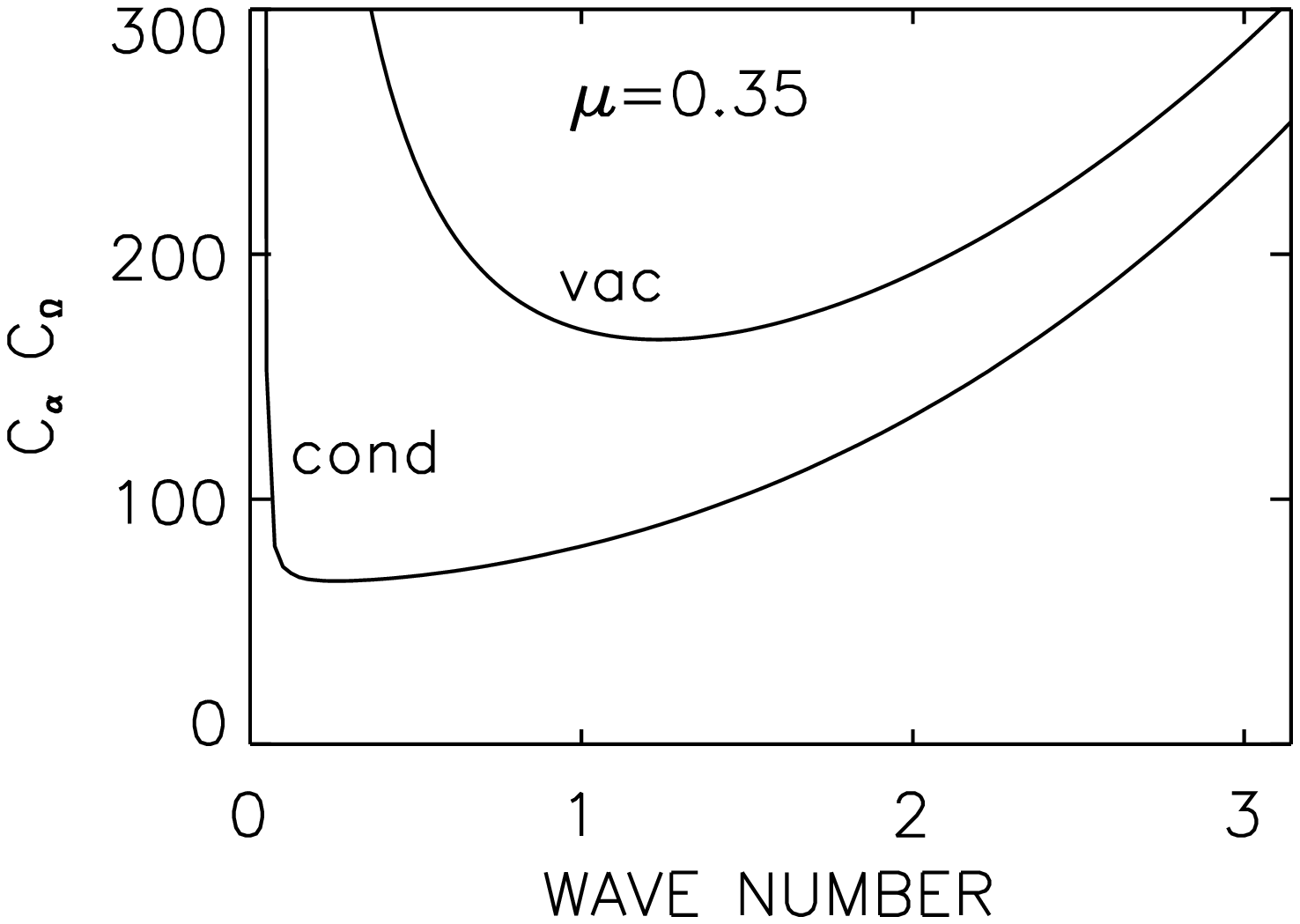}
 \includegraphics[width=0.45\textwidth]{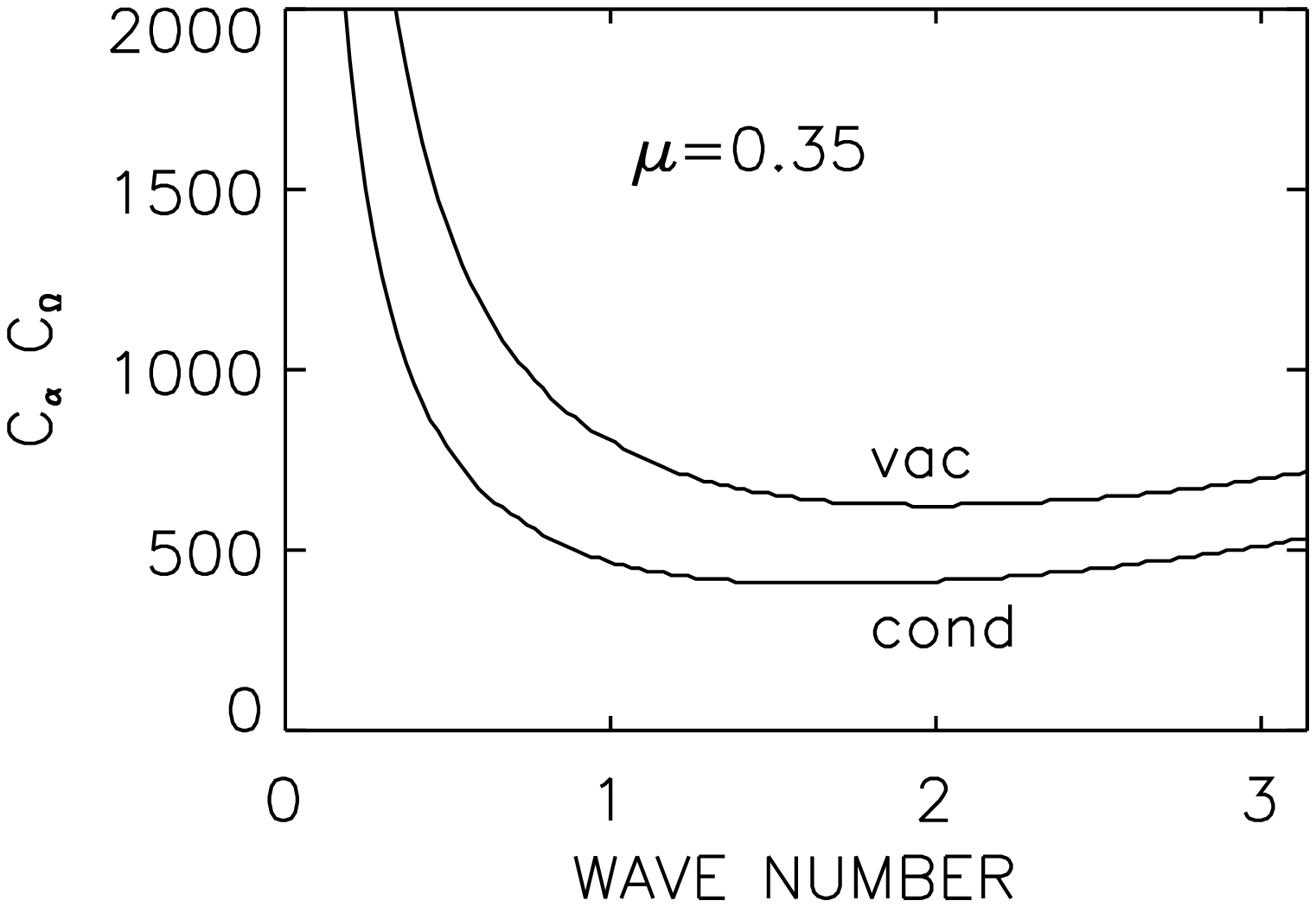}
 }
  \caption{Dynamo number ${\cal D}=C_\alpha C_\Om$ for non-uniform $\alpha$ profiles and quasi-Keplerian rotation law. Top: half sine-type $\alpha$ profile (no sign change between $\Rin$ and $\Rout$), bottom: sine-type $\alpha$ profile (one sign change of $\alpha$ between $\Rin$ and $\Rout$). $C_\Om=1000$,  $\rin=0.5$,  $\mu=0.35$. Insulating (vac) and perfect-conducting (cond)  cylinder walls.}
\label{fig38}
\end{figure}

In order to model zeros of the $\alpha$ effect at the walls the  dynamo model has been modified by use of a half sine-type  profile so that the  $\alpha$ is positive or negative throughout the  gap and the maximum lies in the gap center. The upper panel of Fig. \ref{fig38} shows the resulting  numbers $\cal{D}$ obtained for both sorts of boundary conditions. For vacuum  conditions the eigenvalues are similar  to those for uniform $\alpha$.
The oscillation frequencies for positive $\alpha$ have the opposite sign than those for negative $\alpha$ (see Fig. \ref{fig37}). Most striking, however,  is the behavior of the eigenvalues for the  models with perfect-conducting cylinders. Here the minimum wave number is much smaller than for vacuum boundary conditions. The field geometry becomes more and more two-dimensionally. For $K=0$ the induction  of the differential rotation vanishes and only a non-oscillating $\alpha^2$ dynamo survives.


With Fig. \ref{fig34a} we have offered a     { sinusoidal} $\alpha$ effect  for small $\Pm$  if only one mode  is excited. The  $\alpha$ profile possesses a zero in the gap center and positive as well as negative values along the radius. Whether the inner part of the gap has positive or negative sign only depends on the   mode considered with  $m=1$ or 
$m=-1$.  The critical $C_\alpha$ for the $\alpha^2$ dynamo  must thus  be expected as basically  higher than 3.17. 
For both sorts of sinusoidal  $\alpha$ effect the critical value of  $C_\alpha$ for $\alpha^2$ dynamos  is $C_\alpha = 4.85 $ for conducting boundaries and  $C_\alpha= 4.91 $ for insulating boundaries.  
Because of the small values of the simulated  $C^{\rm sim}_\alpha<1$   an $\alpha^2$ dynamo on basis of the Tayler instability  can therefore not exist. 

For slow rotation the dynamo is stationary while it oscillates for faster rotation. The normalized frequencies are very similar to the numbers in the lower panel of Fig. \ref{fig37}. There is no visible influence of  the form of the radial profile of the $\alpha$ effect.
The lower panel of Fig. \ref{fig38} demonstrates that again the dynamo number $\cal D$ for conducting walls is smaller than for insulating walls.
The minimum value for the model with the vacuum boundary conditions  
 for the sine-type model 
 results as ${\cal D}\approx 700$ while for a quasi-uniform $\alpha$ effect  ${\cal D}\approx 200$ results.
With sufficiently large  $C_\Om$ mean-field dynamos with very small $C_\alpha$ are   always  possible. 
From Eqs. (\ref{sim}) and (\ref{D}) one obtains
\begin{equation}
C^{\rm sim}_\alpha \Rm \gsim {{\cal D}} \left(1+\frac{\eta}{\eta_{\rm T}}\right)
\left(1+\frac{\eta_{\rm T}}{\eta}\right)
\label{sim2}
\end{equation}
as the dynamo condition  where the right-hand side of this relation  always equals or even exceeds   $4 {\cal D}$.  Obviously, with uniform $\alpha$ effect for   $\Rm<\Rm_{\rm crit}$ with $\Rm_{\rm crit}= 4 {\cal D}/C^{\rm sim}_\alpha$ no dynamo excitation is possible.  On the other hand, this relation also means $\Rey\to \infty$ for $\Pm\to 0$ so that always a (small) magnetic Prandtl number exists below which even a hypothetical  dynamo certainly decays. In that sense  our attention is   focused  to models with (say) $\Pm=0.1$. For smaller $\Pm$ dynamo operation is even more complicated. 

Whether  for  larger magnetic Reynolds numbers a dynamo can work or not depends on the details of  the eddy diffusivity theory. The dynamo action is basically suppressed for  $\etaT\gg \eta$ {\em and} also for $\etaT\ll \eta$. If, however,   a saturation of $\etaT/\eta$ for large $\Rm$ and/or $\Lu$ exists then the  dynamo action is always   possible  for the resulting sufficiently  large values of $\Rm$. 
One finds that  mainly the assumptions about      $\etaT/\eta$   decide whether a Tayler dynamo mechanism may work. 

Figure  \ref{fig38} (bottom) also demonstrates the existence of a minimum for the eigenvalue  at $K\simeq 1.5$ which is at least by a factor of two smaller than the vertical wave numbers $k$ of the Tayler instability. For smaller $K$ the shear effect  in the dynamo is too small and for larger $K$ the dissipation is too large.
\section{Discussion and conclusions}\label{Discussion}
The stability of  Taylor-Couette flows with conducting material has been discussed where an axial homogeneous current produces a toroidal magnetic background field. The rotation rates of the cylinders are prescribed;  only solid-body and quasi-Keplerian rotation  laws are here applied. A nonaxisymmetric Tayler instability appears in this model if the rotation is not too fast. 
For given (supercritical) Hartmann number always a maximal Reynolds number $\Remax$ exists above which the fluid is stable. $\Remax$ also depends on the magnetic Prandtl number, $\Remax\propto \Pm^{-0.16}$ for  Kepler rotation and small $\Pm$. The $\Pm$-dependence disappears for \C-type flows, where  the lines of marginal stability in the ($\Ha/\Rey$) plane do not depend on the (small) magnetic Prandtl number.  

The eigenfunctions $\vec u$ and $\vec b$ have been computed by means of the linearized equations. 
The two Fourier modes with $m=1$ and $m=-1$ of the  instability are degenerated  possessing identical critical Hartmann numbers and Reynolds numbers for excitation.  Each of them is helical with opposite values of $\langle \vec{u}\cdot \curl\ {\vec u} \rangle$ and $\langle \vec{b}\cdot \curl\ {\vec b} \rangle$. Both the total sums of the kinetic and magnetic helicity  vanish. It is thus clear that in the mean-field formulation of dynamo theory the $\alpha$ effect vanishes if the two possible modes are coexisting\footnote{Because of the topology of the model also the $\vec{\Om} \times \vec{J}$ term of the turbulence electromotive force vanishes. }.  

One can ask  whether dynamo action is possible if only one of the modes is excited.
Then kinetic and  magnetic helicity exist even without rotation
but  these values are subcritical for dynamo action.  Calculating the axial and the azimuthal components of the electromotive force with  the eigenfunctions of the linearized equation system  the normalized $\alpha$ effect, $C_\alpha$,  results without any free or unknown parameter. Its value never exceeds unity, and the  dependences on the magnetic Prandtl number (for $\Pm\lsim 1$) and the rotation rate are  weak. 
Averaging over the entire container one obtains $C^{\rm sim}_\alpha=O(0.1)$ which is not large enough for the operation of an $\alpha^2$ dynamo. 



However, an $\alpha\Om$ dynamo with $\alpha$ effect {\em and}
differential rotation  may work where -- if the latter is only strong enough -- the $\alpha$ term can even be very weak. 
The resulting  dynamo condition 
\beg
{\Rm} \geq \frac{\cal D}{C^{\rm sim}_\alpha} \left(2+\frac{\etaT}{\eta}  + \frac{\eta}{\etaT}\right)
\label{Rm}
\ende
provides a  lower limit of the  magnetic Reynolds number for dynamo excitation of  $\Rm_{\rm crit}=4 {\cal D}/C_\alpha^{\rm sim}$ as the absolute   minimum of the bracket expression  is 4. From the upper panel of Fig. \ref{fig38} we take ${\cal D}=100$  for perfect-conducting boundaries as the most optimistic eigenvalue of dynamo excitation. On the other hand,  the numerical value $C_\alpha^{\rm sim}$  remains nearly constant for all Reynolds numbers, magnetic Prandtl numbers and shear values (see Fig. \ref{fig35}) hence  one finds  dynamo excitation as only possible for  $\Rm\gsim 2.000$.
This is a large value which  excludes the possibility of related dynamo experiments with liquid metals in the laboratory.  For $\Pm=10^{-5}$ it is  $\Rey_{\rm crit}=4\cdot 10^8$ below which  dynamo excitation is  excluded. Note also that for  $\Pm\to 0$ the critical Reynolds number for dynamo action grows to infinite.

For $\Pm=0.1$ the minimal Reynolds number for dynamo excitation is  $\Rey_{\rm crit}\simeq 20.000$ which needs   Hartmann numbers 
 \beg
\Ha\geq \Ha_{\rm crit}
 \label{Hcrit}
 \ende
with $\Ha_{\rm crit}=207$ for quasi-Keplerian rotation\footnote{For  example: a pinch current of 50 kA   with  $\rin=0.5$ generates  Hartmann numbers of $\Ha=129 $ for GaInSn and $\Ha=406$ for liquid sodium, resp.}.  For smaller Hartmann numbers it is always $\Remax<\Rey_{\rm crit}$ because of the  rotational suppression of the Tayler instability. On the other hand, the critical Hartmann number lies slightly below the Hartmann number $\Ha_1=\S_1/\sqrt{\Pm}$ which after (\ref{diff7}) defines the magnetic field for which $\etaT=\eta$. It is thus clear that close to the Hartmann number $\Ha_{\rm crit}$ an $\alpha\Om$ dynamo may exist. 

 The relation (\ref{Hcrit})  forms a {\em necessary} condition for dynamo excitation, the corresponding minimum Mach number for $\Pm=0.1$ is $\Mm\simeq 30$. The values are  obviously  too large for the  numerical simulation of the magneto-turbulence in Taylor-Couette flows. We have thus to work with simplified models to provide eddy diffusivity values in dependence on the applied (large) Reynolds numbers. 
\begin{figure}
\hskip-1.1cm
\hbox{ 
\includegraphics[width=0.50\textwidth]{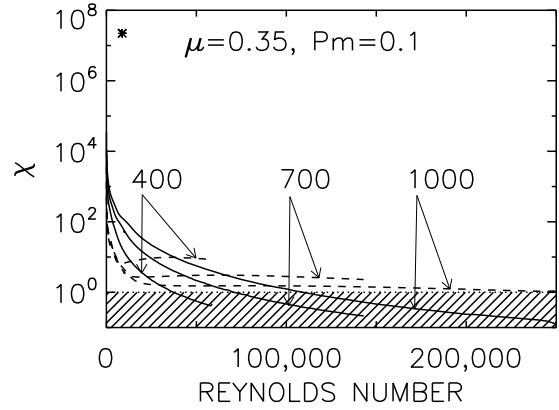}
 }
  \caption{Ratio $\chi$  for magnetic-equilibrated turbulence with ${\cal D}=100$ for quasi-Keplerian rotation and for three different Hartmann numbers (marked).  Each curve ends with its maximal Reynolds number $\Remax$. Solid lines:  $\kappa_ {\rm eq}=1$, dashed lines: $\kappa_ {\rm eq}=0.005$.
    Dynamo self-excitation ($\chi\leq 1$, shaded area) is only   possible for  large Reynolds numbers (i.e. $\Mm\gsim 30$) and for  $\kappa_ {\rm eq}>0.005$.  
    The asterisk represents a model  with $\Pm=10^{-5}$ for $\Ha=50$ and $\Remax=8812$.  
$\rin=0.5$, $\mu_B=2$, $\mu=0.35$, $\Pm=0.1$. Perfect-conducting cylinder walls.}
\label{fig43}
\end{figure}
Let us first numerically study the  dynamo condition in the form
\beg
 \chi =\frac{ {\cal D} }{ C^{\rm sim}_\alpha\  \Rm}\left(2+\frac{\etaT}{\eta}  + \frac{\eta}{\etaT}\right) \leq 1.
\label{chi}
\ende
As the numerical value of $C_\alpha$ does hardly depend on the Reynolds number and the magnetic Prandtl number,   
it is thus clear that mainly the ratio $\etaT/\eta$ decides about  dynamo excitation or not. Both  large   and also small values of $\etaT/\eta $  hinder the dynamo excitation. If $\etaT/\eta $  is independent of $\Rm$ and/or $\Lu$ -- or if by saturation a maximal value exists -- then  always a (large) $\Rm\geq \Rm_{\rm crit}$ exists for which $\chi$ becomes small enough so that the dynamo condition is fulfilled. For $\Rm<\Rm_{\rm crit}$  the condition (\ref{chi}) can never be fulfilled. 
\begin{figure}
\centering
\includegraphics[width=0.5\textwidth]{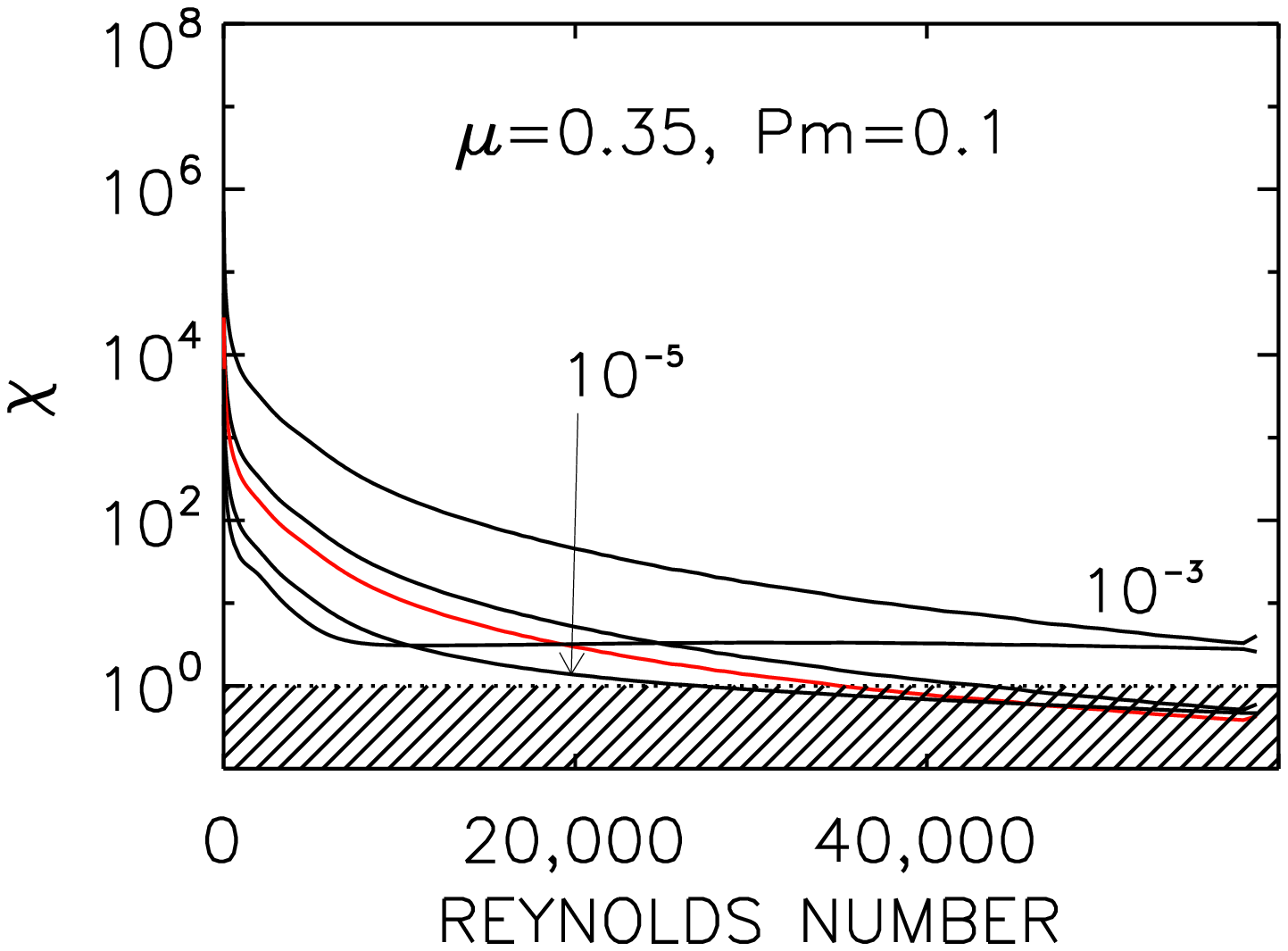}
\includegraphics[width=0.5\textwidth]{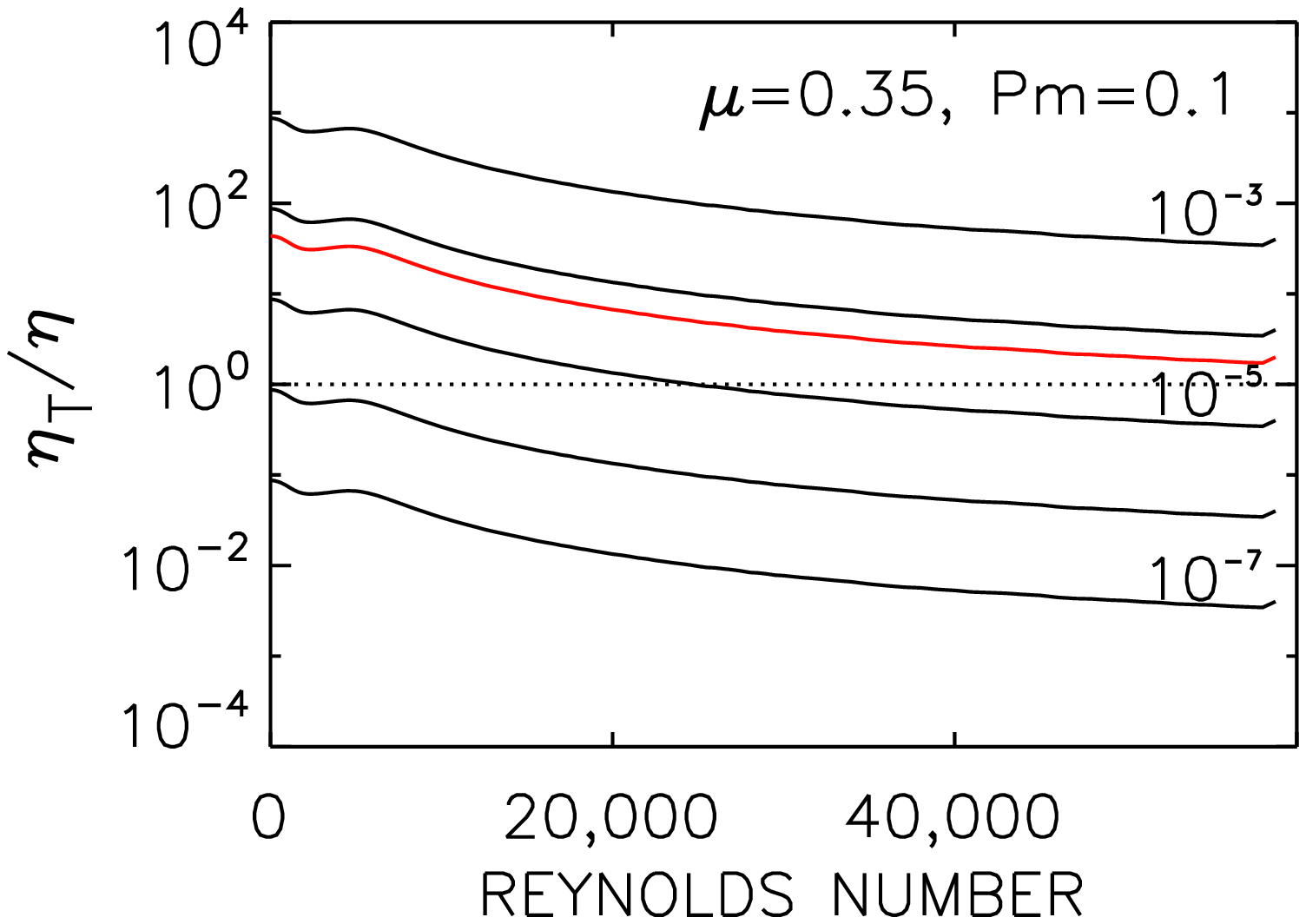}
 \caption{Similar as Fig. \ref{fig43} but  for the nonlinear turbulence model (\ref{diff6})   for   fixed Hartmann number $\Ha=400$ for all   $\Rey\leq \Rey_{\rm max}$. 
 $\kappa$ (marked) is uniform between $\Rin$ and $\Rout$. Top panel: ratio $\chi$, bottom panel: ratio $\etaT/\eta$. The horizontal dotted line gives $\etaT=\eta$.  The red lines belong to $\kappa=5\cdot 10^{-5}$. ${\cal D}=100$, $\rin=0.5$, $\Pm=0.1$, $\mu=0.35$. Perfect-conducting cylinder walls. }
\label{fig44a}
\end{figure}
\begin{figure}
\centering
\includegraphics[width=0.5\textwidth]{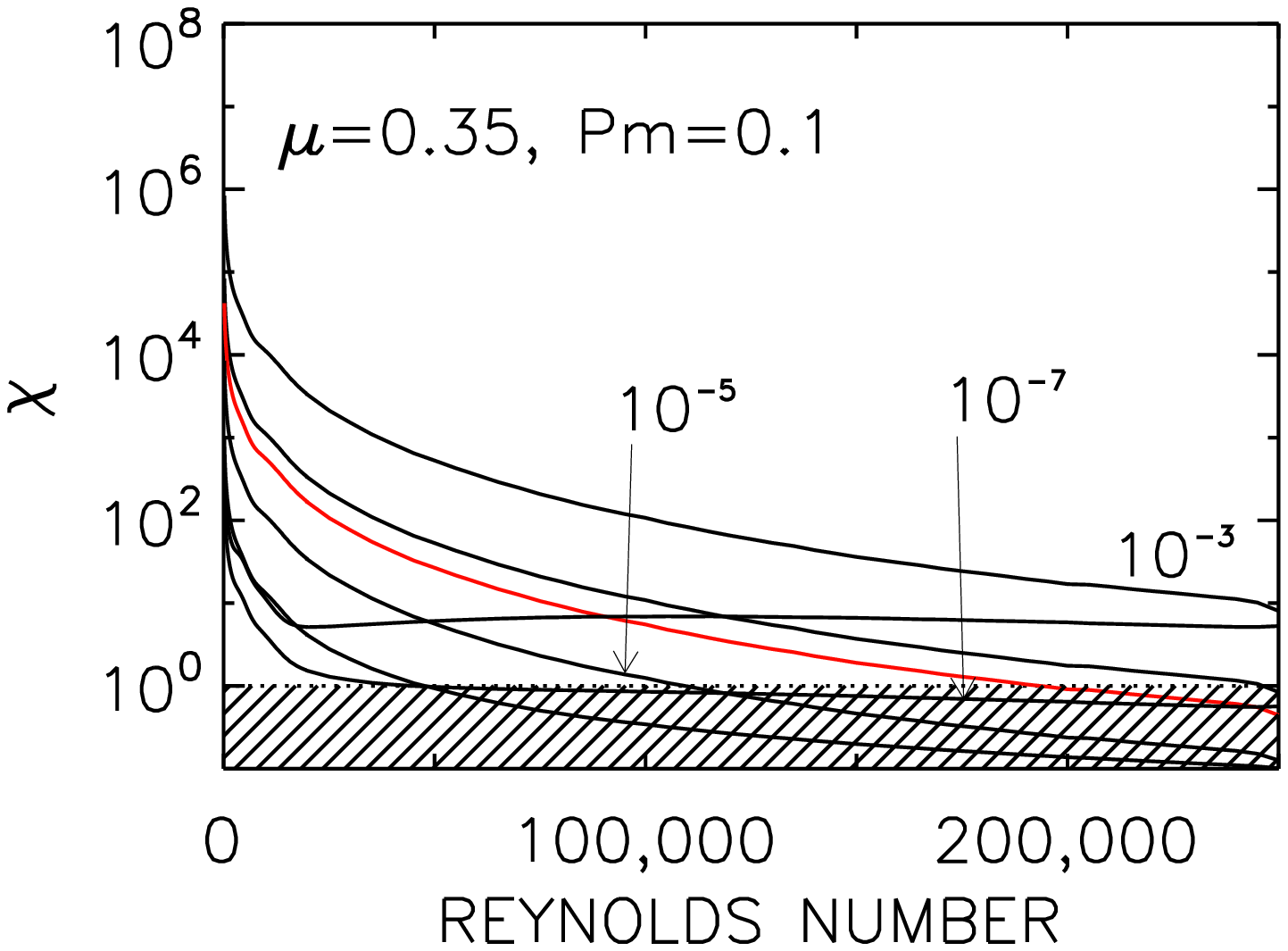}
\includegraphics[width=0.5\textwidth]{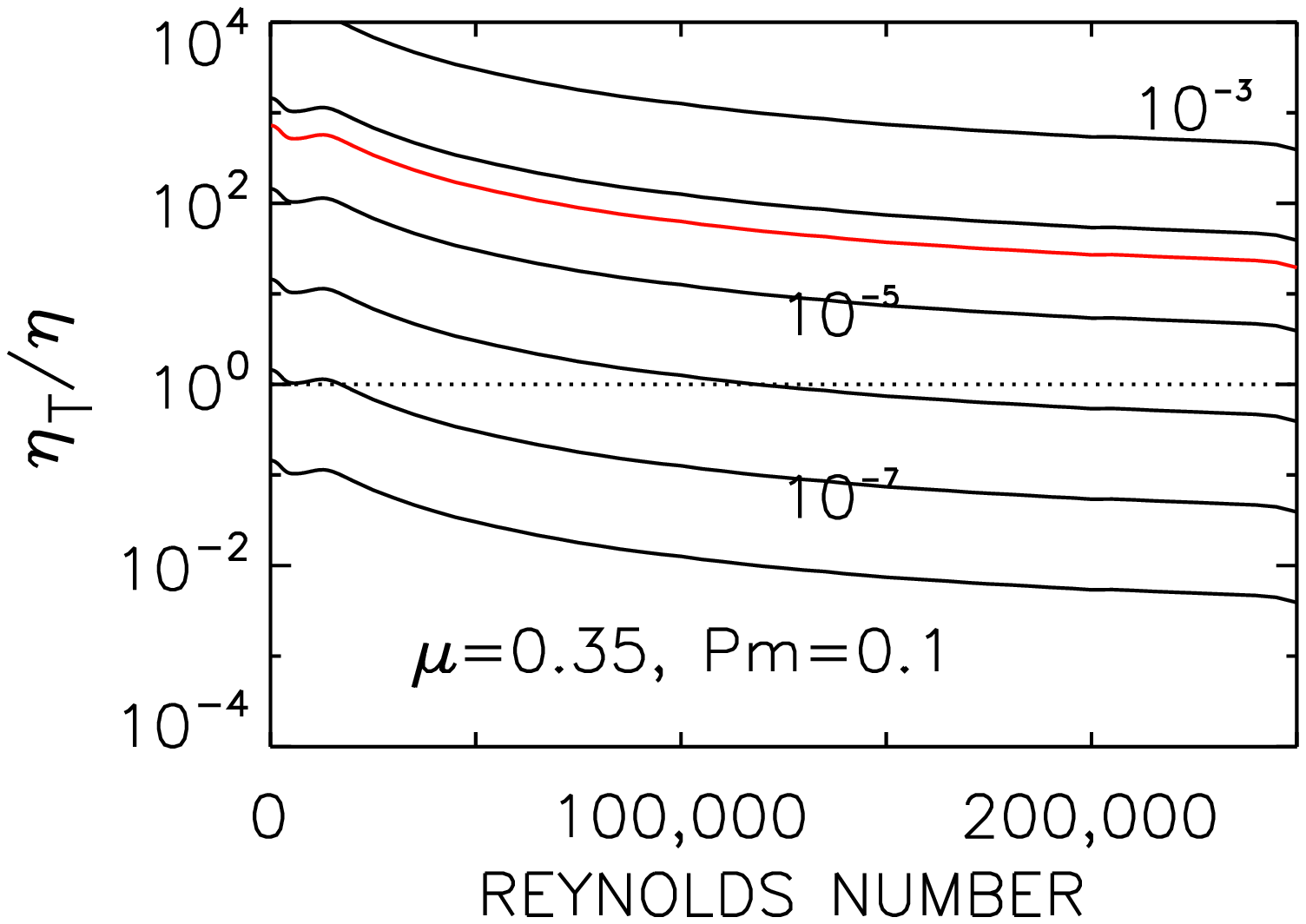}
 \caption{Similar as Fig. \ref{fig44a} but  for  $\Ha=1000$ and all   $\Rey\leq \Rey_{\rm max}$. 
}
\label{fig44b}
\end{figure}
That  the possible existence of  dynamo action depends on the details about the instability-induced diffusivity bases on the two features of the Tayler instability that i) the simulated  $C^{\rm sim}_\alpha$   is nearly constant for all magnetic fields and rotation rates and that  ii) the rotation rate has a maximal value which cannot be exceeded unless the instability decays.  
 
 If the $\etaT/\eta $ grows  {\em linearly} with the Lundquist number $\Lu$  then the  numerical values of the  parameters (including the magnetic Mach number) decide whether the dynamo condition is fulfilled or not. Figure \ref{fig43} demonstrates for the quasi-linear model defined by Eq.  (\ref{diff4}) that  for $\Ha\geq 400$ dynamo solutions always exist (if $\kappa_{\rm eq}=1$). The $\chi$'s take their minimum for  the largest Reynolds numbers $\Remax$. Only   rapidly rotating  containers can thus be dynamo-active. Here the $\Remax$ (above which  the Tayler instability decays) is large enough to fulfill the  condition (\ref{chi}). The magnetic Mach number must exceed a critical value (here $\Mm\simeq 10$),  which  for quasi-Keplerian rotation and the chosen  magnetic Prandtl number (here $\Pm=0.1$) is always possible. 
 
It is $\etaT>\eta$  along all curves with $\kappa_ {\rm eq}=1$ in Fig. \ref{fig43}.  Because of the rotational suppression of the correlation time  $\tau_{\rm corr }$ shown in Fig. \ref{fig36a} the fast-rotation parts of the curves possess much smaller eddy diffusivities than their slow-rotating parts. Only the fast-rotating parts are  located in the dashed areas of dynamo self-excitation. It is also obvious that models with $\kappa_ {\rm eq}<1$ are even more dynamo-active than the plotted examples with $\kappa_ {\rm eq}=1$. The effective Mach number in this case is increased. For too small $\kappa_ {\rm eq}$, however, the eddy diffusivity becomes smaller than the microscopic one and the  bracket in (\ref{Rm}) becomes large enough to suppress the dynamo. The smallest possible $\kappa_ {\rm eq}$ for dynamo self-excitation of the examples shown in Fig. \ref{fig43} is 0.005.  Obviously, the dynamo-instability of this model is rather robust.


 We shall  proceed with  a  diffusivity model for which the diffusivity grows  faster  with the magnetic field.  In this case the dynamo condition (\ref{chi})  should also have  limits where the magnetic diffusivity is too large for dynamo excitation. The  highly nonlinear  model  (\ref{diff6})  is  applied.  We note that along the ``vertical slices'' in the ($\Ha/\Rey$) plane the growth rates of the instability are positive  for  $\Rey<\Rey_{\rm max}$. We note that their  maxima for fixed $\Ha$  exist at  small Reynolds numbers  shortly above the horizontal axis where $\alpha\Om$ dynamos do certainly not exist.
 
Figure \ref{fig44a} gives  the function $\chi$ defined by (\ref{chi}) for   $\Ha=400$ and various  $\kappa$. The   Hartmann number exceeds the minimum value (\ref{Hcrit}) so that dynamo solutions  are not basically excluded. 
The top panel of this figure  shows the excitation conditions 
while   the bottom panel gives the corresponding diffusivity values $\etaT/\eta$. We find  dynamo excitation with the lowest Reynolds number  for  $\kappa=10^{-5}$. The eddy diffusivity for  this $\kappa$ value is of order unity. The larger value $\kappa=5\cdot 10^{-5}$ 
known from Section \ref{Eddy}
for  for $\Ha_1=316$ 
also leads to  dynamo excitation with somewhat higher value of $\etaT/\eta$. For the smaller value $\kappa=10^{-6}$ the eddy diffusivity is too small for dynamo excitation. 
On the other hand,  dynamo action  does not appear for   $\kappa>10^{-4}$.  The eddy diffusivity for lower or higher $\kappa$ values  proves be too low or too high  for the  $\alpha\Om$ dynamo mechanism.
Because of the existence of the upper limits $\Remax$ of the Reynolds numbers the curves for $\kappa>10^{-4}$  in  Fig. \ref{fig44a} (top) are not long enough to reach the shaded area. 

Similar  data  for  $\Ha=1000$   are given by Fig. \ref{fig44b}.
 One finds  the excitation with the lowest Reynolds number  for the smaller value  $\kappa=10^{-6}$. Again the specific  eddy diffusivity for  this $\kappa$ value is {of order unity}. The larger value $\kappa=5\cdot 10^{-5}$ (red line) moved upwards in comparison to Fig. \ref{fig44a}; it will finally  leave the shaded dynamo excitation area for $\Ha> 1000$ because of the increase of  $\etaT/\eta$ for increasing Lundquist number.
One also finds  that for $\kappa=10^{-8}$ the eddy diffusivity is too small for dynamo excitation
and for $\kappa=10^{-3}$ it is too large. 
Just for $\etaT=\eta$ the bracket in the dynamo condition (\ref{chi}) takes its lowest value hence the self-excitation is easiest.
Again it is demonstrated  by the upper panel of Fig. \ref{fig44b} that  because of the existence of the  maximal  Reynolds number (here $\Remax\simeq 250.000$) the curves for too small or too large $\kappa$  are simply not long enough for dynamo excitation.

That the values of $C^{\rm sim}_{\alpha}$ are almost uniform  in the entire instability domain does not mean that the $\alpha$ effect and the eddy diffusivity are almost constant. We only know that their ratio is almost constant. For differentially rotating fluids an $\alpha\Om$ dynamo could exist but we find that  the maximal possible rotation rates (defined by $\Remax$) can easily  be  too slow to maintain the dynamo.  On the other hand, the existence of large-scale dynamo action for solid-body rotation can  be excluded with ease.


The influence of the magnetic Prandtl number on the dynamo excitation is still an open question. We also  note that the $\chi$ values drastically grow for smaller $\Pm$.  The asterisk in Fig. \ref{fig43} represents an isolated calculation of a model with quasi-Keplerian rotation for $\Pm=10^{-5}$  demonstrating the massive stabilization  by small magnetic Prandtl numbers.  As an example, the critical  Reynolds number $\Rey_{\rm crit}$  allowing  dynamo excitation for $\Pm=10^{-5}$ is $2\cdot 10^8$, which is far beyond our numerical limitations.  
The following argument goes in a similar direction.  It is obvious that  the $\Pm$-dependence of the eddy-diffusivity determines the $\Pm$-dependence in the dynamo condition (\ref{chi}). With numerical simulations \cite{GR09} demonstrated that $\etaT/\eta$ for given Hartmann number does not depend on the magnetic Prandtl number, hence $\etaT/\eta\propto \Ha$, or similarly, $\etaT/\eta\propto \Lu/\sqrt{\Pm}$. For large enough $\etaT/\eta$ the dynamo condition (\ref{chi}) turns into $\chi\propto \Mm^{-1}\Pm^{-1/2}$ which for small $\Pm$ grows to large values.

There is thus an   indication that small $\Pm$  suppresses the dynamo-instability  of the Tayler pattern. The deeper reason of this phenomenon  may be that the dynamo condition (\ref{Rm})  basically scales with the magnetic Reynolds number while the rotational quenching of the Tayler instability  scales with the ordinary Reynolds number which both strongly differ for small magnetic Prandtl numbers. 

\section{acknowledgment}
Frank Stefani from the Helmholtz-Zentrum Dresden-Rossendorf is acknowledged for many discussions and several  critical readings of the manuscript.

\bibliographystyle{mn2e}
\bibliography{superamri}

\end{document}